\documentclass[AMA,STIX1COL]{WileyNJD-v2}

\usepackage[utf8]{inputenc}
\usepackage{moreverb}
\usepackage{amsmath}
\usepackage{bm}
\usepackage{nicefrac}
\usepackage{gensymb}
\usepackage[all]{nowidow}
\usepackage{algorithm}
\usepackage[format=hang]{subcaption}
\usepackage{hyperref}
\usepackage[capitalise,noabbrev]{cleveref}
\usepackage{siunitx}
\usepackage{placeins}
\usepackage{xcolor}
\usepackage{pgfplotstable} 
\usepackage{physics}
\usepackage{comment}

\usepackage{tikz}
\usepackage{pgfplots}
\pgfplotsset{compat=1.5.1}
\usetikzlibrary{calc}
\usetikzlibrary{arrows}
\usetikzlibrary{patterns}

\newcommand*\diff{\mathop{}\!\mathrm{d}}
\newcommand*{\ddt}[1]{\dfrac{\text{d} #1}{\diff t}}
\newcommand*{\vmmem}[0]{\emph{VM-MEM-LBM}}
\newcommand*{\mem}[0]{\emph{MEM-LBM}}
\newcommand*{\thyd}[0]{\bm{T}_p^\mathit{hyd}}
\newcommand*{\fhyd}[0]{\bm{F}_p^\mathit{hyd}}
\newcommand*{\fg}[0]{\bm{F}_p^\mathit{g}}
\newcommand*{\fv}[0]{\bm{F}_p^\mathit{v}}
\newcommand*{\fp}[0]{\bm{F}_p}
\newcommand*{\tp}[0]{\bm{T}_p}
\newcommand*{\tv}[0]{\bm{T}_p^\mathit{v}}
\newcommand*{\pip}[0]{\pi_p}
\newcommand*{\cv}[0]{C_v}
\newcommand*{\cvomega}[0]{C_{v,\omega}}
\newcommand*{\Ga}[0]{\mathrm{Ga}}
\newcommand*{\St}[0]{\mathrm{St}}
\newcommand*{\ZD}[0]{Zhou \& Dušek}
\newcommand*{\AM}[0]{Auguste \& Magnaudet}

\newcommand\BibTeX{{\rmfamily B\kern-.05em \textsc{i\kern-.025em b}\kern-.08em
T\kern-.1667em\lower.7ex\hbox{E}\kern-.125emX}}

\articletype{Article Type}

\received{<day> <Month>, <year>}
\revised{<day> <Month>, <year>}
\accepted{<day> <Month>, <year>}

\newcommand{\walberla}{{\mdseries \scshape waLBerla}}

\begin{document}

\title{Coupling fully resolved light particles with the Lattice Boltzmann method on adaptively refined grids}

\author[1]{Lukas Werner*}
\author[1]{Christoph Rettinger}
\author[1,2]{Ulrich Rüde}

\authormark{Lukas Werner \textsc{et al}}

\address[1]{\orgdiv{Chair for System Simulation}, \orgname{Friedrich--Alexander--Universität Erlangen--Nürnberg}, \orgaddress{Cauerstraße 11, 91058 Erlangen, \country{Germany}}}
\address[2]{\orgname{CERFACS}, \orgaddress{42 Avenue Gaspard Coriolis, 31057 Toulouse Cedex 1, \country{France}}}

\corres{*Lukas Werner, Cauerstraße 11, 91058 Erlangen, \email{lks.werner@fau.de}}

\presentaddress{\orgdiv{Chair for System Simulation}, \orgname{Friedrich--Alexander--Universität Erlangen--Nürnberg}, \orgaddress{Cauerstraße 11, 91058 Erlangen, \country{Germany}}}

\abstract[Abstract]{
The simulation of geometrically resolved rigid particles in a fluid relies on
coupling algorithms to transfer momentum both ways between the particles and the fluid.
In this article, the fluid flow is modeled with a parallel Lattice Boltzmann method 
using adaptive grid refinement to improve numerical efficiency.
The coupling with the particles is realized with the momentum exchange method.
When implemented in plain form, instabilities may arise in the coupling when the particles are lighter than the fluid.
The algorithm can be stabilized with a virtual mass correction specifically developed for the 
Lattice Boltzmann method.
The method is analyzed for a wide set of physically relevant regimes,
varying independently the body-to-fluid density ratio and the relative magnitude of inertial and viscous effects.
These studies of a single rising particle exhibit periodic regimes of particle motion as well as chaotic behavior, as previously reported in the literature.
The new method is carefully compared with available experimental and numerical results.
This serves to validate the presented new coupled Lattice Boltzmann method and additionally it
leads to new physical insight for some of the parameter settings.
}


\maketitle

\section{Introduction}

Light particles, here understood as particles with a solid-to-fluid density ratio $\pip = \nicefrac{\rho_p}{\rho_f} < 1$,
appear in various applications and situations. 
During ascension in an otherwise resting fluid they exhibit a multitude of movement patterns often strongly deviating from straight paths.
This depends on shape, density ratio, and surrounding fluid properties.
As a matter of fact, strictly straight paths only occur in a comparatively small parameter range.
This makes way to consider the implications of such non-regular paths in practical settings.
As an environmental 
problem, floatsam in the form of microplastics with a density slightly 
lower than 
water may harm the biosphere in oceans \cite{Andrady2011,Wright2013,Cole2011} 
and lakes \cite{Driedger2015}. 
Here the density ratio ranges 
from $\pip \approx 0.9$ for polypropylene 
to $\pip \approx 0.01$ in the case of expanded polystyrene \cite{Driedger2015}.
Furthermore, in a natural setting, rising bubbles and particles in oceans 
support the formation of warmer regions with stronger temperature gradients \cite{Thorpe1988,Weller1990}.
Facilitating the dispersion of fluid and matter, the turbulent structures induced by moving bubbles of low density ratios in a mixture is also of interest in engineering applications \cite{Almeras2015,Mathai2018a,Mathai2020}.
Two examples are fuel sprays in combustion engines and bubbly flows in industrial reaction catalysis \cite{Bourgoin2014,Mathai2018}.
In aerodynamics the effects are observed for meteorology balloons \cite{Murrow1964,Scoggins1964}, and in the context of autorotation of individual objects \cite{Lugt1983}.

Above examples typically involve a huge number of particles, whereas even the movement of a single spherical particle exhibits complex behaviors.
A light, spherical body submerged in water under the sole influence of gravity may
either rise in a straight line, move obliquely, enter a regular or rather chaotic zig-zagging movement.
As reported by Jenny \etal\cite{Jenny2004}, this depends on the density ratio $\pip$
and the Galileo number $\Ga$, which describes the ratio between buoyant and viscous forces.
They used numerical simulation to study the particle paths, varying these two parameters.
In the following, Veldhuis \& Biesheuvel \cite{Biesheuvel2007}, Horowitz \& Williamson \cite{Horowitz2010} and Ostmann, Chaves \& Brucker \cite{Ostmann2017}, 
among others, performed laboratory experiments to obtain physical comparison results.
To address the found contradictions, Zhou \& Dušek \cite{Zhou2015} and Auguste \& Magnaudet \cite{Auguste2018} performed an extensive set of detailed numerical simulations. 
Albeit very efficient for such specific setups, these recent simulation approaches are usually not extensible to multiple particles or arbitrary domains due to the usage of spectral methods, making them only suitable for a single freely moving sphere.

Such restrictions are not present for other methods in computational fluid dynamics (CFD), such as finite volume methods \cite{versteeg2007introduction,moukalled2016finite} or the lattice Boltzmann method \cite{Chen1998,Aidun2010} (LBM).
Many of the previously mentioned scenarios involving bubbly flows
can be 
approximated by modeling those submerged small bubbles as rigid spheres in a fluid due to their low 
Eötvös number \cite{Clift2008}. 
This enables the usage of a Lagrangian description for the solid phase by employing discrete particle simulation.
In combination with fluid-particle coupling that enables geometrically fully resolved simulations of the flow around the particle and the fluid-solid interactions, predictive studies of a large variety of particulate systems become possible~ \cite{Rettinger2017Riverbed,vowinckel2019,peng2019,benseghier2020}.
In this work, we focus on the LBM which is a relatively recent Eulerian computational method to simulate fluids.
The LBM is well suited for large scale simulations and massively parallel execution \cite{Goetz2010,hasert2014,Bauer2020}.

The fully coupled interaction between the solid Lagrangian and the fluid Eulerian phase is achieved by the momentum exchange method (\mem{})~\cite{Ladd1994,Aidun1998,Rettinger2017}.
However, the \mem{} on its own is only partially applicable for simulations of light particles, as intended in the present work. 
It is known to suffer from instabilities when density ratios approach zero.
The stability condition of the original coupling by Ladd~\etal{} \cite{Ladd1994_2} was given 
as $\pip d_p / \Delta x > 20$, where $d_p / \Delta x$ is the numerical resolution in terms of cells per particle diameter.
With the improvements by Aidun~\etal{} \cite{Aidun1998}, density ratios close to one could be realized. 

This issue of numerical instability at those circumstances is not restricted to the \mem{} and reported for many CFD methods that apply an explicit fluid-particle coupling.
For the popular immersed boundary method\cite{Uhlmann2005} (IBM), e.g.,
the lower bound initially resided at $\pip = 1.2$.
Improvements to the imposition of boundary conditions on the interface and an additional integration step for the artificial flow field inside the particles
allowed stable results at $\pip > 0.3$
\cite{Kempe2012,Breugem2012}.
To gain a fully stable solution at arbitrary density ratios, much more costly implicit schemes were proposed \cite{Inamuro2004,Apte2013,banks2018}.

The reason for the instabilities in explicit coupling schemes is attributed to the particles' mass being exceeded by its so-called added mass, i.e.~the fluid attached to the moving particle.
In an explicit coupling scenario, this mass cannot be properly accounted for, which results in excessive accelerations on startup that lead to oscillations\cite{Jenny2004a,Hu2001}.
In an effort to stabilize explicit schemes at density ratios as low as $\pip = 0.001$,
Schwarz et al.~investigated the cause of stability issues in the IBM \cite{Schwarz2015}.
By introducing an auxiliary virtual force and a virtually increased mass of purely numerical nature to the particle, stable simulations of light particles could be obtained.
This scheme was termed the ``virtual mass approach'' by the authors.

In this paper, we aim to improve the existing \mem{} and extend it to simulations of particles with very low density ratios.
To this end, we follow and improve the approach by Schwarz \etal, resulting in a \mem{} enhanced by virtual mass (\vmmem). 
This will allow for predictive numerical studies of such setups.
As an illustrative example and to demonstrate the applicability of our approach, we investigate the movement of a single rising sphere in an unbounded domain.
In an effort to validate existing results from literature, demonstrate the capabilities of the \vmmem{} and enabling a scalable simulation of submerged light spheres, we apply the LBM to this problem.
To our best knowledge, our article is the first to apply the LBM in a wide range of parameter settings of this situation.

Improving the efficiency of these simulations, an adaptive grid refinement technique is employed. 
Since domains with a height in the order of 100 times of the sphere diameter are necessary, many regions of the fluid remain stale during the motion of the sphere, requiring only coarse resolutions.
On the other hand, increasing turbulences and strong velocity gradients especially at the fluid-solid interface require a high resolution to yield satisfying results.
Thus, a grid, which locally adapts its spacing, is desirable.

The paper is structured as follows. 
In \cref{sec:numerical_methods}, the \mem{} as used for the simulations throughout this paper is outlined.
\cref{sec:extension_light_particles} starts with a numerical study to investigate and discuss the lower density ratio limits of the \mem. 
Next, this method is enhanced by the virtual mass approach, resulting in the \vmmem. 
Efficiency improvements by employing adaptive grid refinement and criteria for determining the need of a change in resolution are explained in \cref{sec:agr}.
\cref{sec:numerical_study} presents an extensive study of the movement trajectories and vortices of rising spheres to demonstrate the practical use of \mem{} and \vmmem.
We conclude with a summary in \cref{sec:conclusion}.

\section{Numerical method}
\label{sec:numerical_methods}
The here employed CFD method is based on the recently developed approach presented and validated in Ref.~\citenum{Rettinger2020b}.
For completeness, we outline the main aspects of the fluid flow simulation and the fluid-particle coupling in the following and refer to this work for a more detailed description.
All presented methods can be found in the open-source repository\footnote{\url{www.walberla.net}} of the \walberla{} framework~\cite{Bauer2020}.

\subsection{Lattice Boltzmann Method \label{sec:lbm}}

The lattice Boltzmann method (LBM) 
is based on an Eulerian representation of the fluid field by mesoscopic kinetic equations,
that fulfill the macroscopic Navier-Stokes equations~\cite{Chen1998, krueger2017}. 
It computes the temporal 
evolution of particle distribution functions (PDFs) $f_q$ 
in the cells of a Cartesian lattice 
using a discrete solution of the Boltzmann equation. 
The LBM is commonly described using the D$d$Q$q$ notation with $d$ spatial dimensions and $q$ discrete velocities $\bm{c}_q$~\cite{Qian1992}.
A common choice for three dimensional simulations is D$3$Q$19$ which will be employed throughout this paper.

Being an explicit numerical scheme in time, the equation to advance the PDFs of a cell $\bm{x}$ by one time step of size $\Delta t$ is given by
\begin{equation}
	f_q(\bm{x}+\bm{c}_q\Delta t, t+\Delta t) = \tilde{f}_q(\bm{x}, t) = f_q(\bm{x}, t) - C_q (f_1(\bm{x},t), ..., f_{19}(\bm{x}, t)) \text{,}
\end{equation}
where $C$ corresponds to a general collision operator. 
It is applied during the collision step, i.e. the right part of the equation, to update the local PDFs in each cell $\bm{x}$ inside the domain. 
The left part denotes the streaming step of the LBM, which constitutes distribution of the post-collision PDFs $\tilde{f}_q(\bm{x}, t)$ to the neighboring cells.

The local macroscopic fluid density $\rho_f$ and velocity $\bm{u}_f$ are given as moments of the distribution functions~\cite{Chen1998}:
\begin{equation}
	\rho_f(\bm{x}, t) = \sum_q f_q(\bm{x}, t),\quad
	\rho_0 \bm{u}_f(\bm{x}, t) = \sum_q f_q(\bm{x}, t) \bm{c}_q \text{,} 
	\label{eq:lbm_uf}
\end{equation}
where $\rho_0$ is a constant average density, that is introduced to reduce inherent compressibility effects~\cite{he1997}.

In LBM, quantities are commonly formulated in so-called lattice units, such that the cell size $\Delta x = 1$, time step size $\Delta t = 1$ and $\rho_0 = 1$.

In the multiple-relaxation-time (MRT) formulation, the collision operator $C$ is written using a diagonal relaxation matrix $\bm{S}$, containing the relaxation factors, and a corresponding matrix $\bm{M}$, which linearly transforms the distribution functions to the moment space, where the collision is carried out~\cite{DHumieres2002}. The collision operator is thus given as
\begin{equation}
	C = \bm{M}^{-1} \bm{S} \bm{M} \text{.}
\end{equation}

In particular, we here employ the specific MRT model from Ref.~\citenum{Rettinger2020b}, which uses the transformation matrix $\bm{M}$ given in Ref.~\citenum{duenweg2007} and the relaxation matrix
\begin{equation}
	\bm{S} = \operatorname{diag}\left( 0,\ 0,\ 0,\ 0, \  s_b, \  s_b, \  s_\nu^-, \  s_\nu^-, \  s_\nu^-, \  s_\nu, \  s_\nu, \  s_\nu, \  s_\nu, \  s_\nu, \  s_\nu, \  s_\nu, \  s_\nu^-, \  s_\nu^-, \  s_\nu^-\right) \text{.}
\end{equation}
Via its three parameters, it allows for explicit control over the kinematic fluid viscosity $\nu_f$ and the bulk viscosity $\nu_b$, given as 
\begin{equation}
\nu_f = c_s^2 \Delta t\left(\tfrac{1}{s_{\nu}}-\tfrac{1}{2}\right),\quad \nu_b = \tfrac{2}{3} c_s^2 \Delta t\left(\tfrac{1}{s_b}-\tfrac{1}{2}\right),
\end{equation}
with the lattice speed of sound $c_s$,
and provides an accurate representation of boundaries~\cite{Rettinger2020b}.
Parameterization can conveniently be done by introducing the so-called ``magic'' parameter $\Lambda$~\cite{Ginzburg2008}, which we set to $\nicefrac{3}{16}$, and a bulk factor $\Lambda_b$~\cite{khirevich2015}
\begin{equation}
\Lambda = \left(\frac{1}{s_\nu} - \frac{1}{2}\right)\left(\frac{1}{s_\nu^-} - \frac{1}{2}\right),\quad \Lambda_b = \frac{\left(\tfrac{1}{s_b}-\tfrac{1}{2}\right)}{\left(\tfrac{1}{s_\nu}-\tfrac{1}{2}\right)}.
\end{equation}
Choosing $\Lambda_b = 1$ results in the two-relaxation-time (TRT) collision operator~\cite{Ginzburg2008}.
Unless stated otherwise, we employ $\Lambda_b = 100$.

Throughout the paper, we consistently make use of the commonly applied lattice unit system, given by the cell size $\Delta x=1$, the time step size $\Delta t =1$, $c_s = 1/\sqrt{3}$, and $\rho_0=1$.

\subsection{Rigid Particle Motion\label{sec:rigid_particle_motion}}

The 
Lagrangian dynamics of a spherical rigid particle is governed by the temporal evolution of its position 
$\bm{x}_p$, translational velocity $\bm{u}_p$ and angular velocity $\bm{\omega}_p$, 
and is 
described by
\begin{subequations}
	\begin{align}
		\ddt{\bm{x}_p} & = \bm{u}_p \text{,} \label{eq:num_methods_particle_translational_basic} \\
		m_p \ddt{\bm{u}_p} & = \fp \label{eq:num_methods_particle_velocity} \text{,} \\
		I_p \ddt{\bm{\omega}_p} & = \tp \label{eq:num_methods_particle_angular_basic} \text{,}
	\end{align}
\end{subequations}
where $m_p = \rho_p V_p$ is the particle's mass, with the particle density $\rho_p$ and its volume $V_p$, and 
$I_p$ corresponds to its moment of inertia.
Gravitation and buoyancy contributions are given as 
$\fg = (\rho_p - \rho_f) \bm{g} V_p$, 
with the gravitational acceleration $\bm{g}$.
Together with the fluid-particle interaction force
$\fhyd$, 
they are the total force acting on the particle, 
$\fp = \fg + \fhyd$.
Accordingly, the particle's torque is 
$\tp = \thyd$.

Temporal integration of \cref{eq:num_methods_particle_translational_basic,eq:num_methods_particle_velocity,eq:num_methods_particle_angular_basic} is achieved by explicit time stepping using a Velocity-Verlet scheme. 
The explicit update formulas are expressed as
\cite{Preclik2015,Rettinger2020b}
\begin{subequations}
	\begin{align}
	\bm{x}_p(t + \Delta t) & = \bm{x}_p(t) + \Delta t_p \bm{u}_p(t) + \dfrac{\Delta t^2}{2 m_p} \fp(t) \text{,} \\
	\bm{u}_p (t + \Delta t) & = \bm{u}_p(t) + \dfrac{\Delta t}{2 m_p}\left(\fp(t) + \fp(t + \Delta t)\right) \text{,} \\
	\bm{\omega}_p (t + \Delta t) & = \bm{\omega}_p(t) + \dfrac{\Delta t}{2 I_p}\left(\tp(t) + \tp(t + \Delta t)\right) \text{.}
	\end{align}
\end{subequations}
The force $\fp(t + \Delta t)$ and torque $\tp(t + \Delta t)$ are computed with the already updated position.
Linear acceleration $\bm{a}_p(t + \Delta t)$ and angular acceleration $\bm{\beta}_p(t + \Delta t)$ are directly given via $\fp(t + \Delta t)$ and $\tp(t + \Delta t)$ as
\begin{subequations}
	\begin{align}
	\bm{a}_p(t + \Delta t) & = \dfrac{\fp(t+\Delta t)}{m_p} \text{,} \\
	\bm{\beta}_p(t + \Delta t) & = \dfrac{\tp(t+\Delta t)}{I_p} \text{.}
	\end{align}
\end{subequations}
Rotation is not explicitly accounted for due to the focus on spherical particles.
Since inter-particle collisions are not regarded in this work, we do not apply sub-cycling of the particle part and thus make use of the same time step size $\Delta t$ here as in the LBM~\cite{Rettinger2020b}.

\subsection{Fluid-Particle Coupling}

The 
coupling between the fluid and solid phase is accomplished via the momentum exchange method (\mem{}) \cite{Ladd1994,Aidun1998} which has successfully been applied to a variety of particulate flow simulations~\cite{Goetz2010,Rettinger2017,Rettinger2020b}.
The core idea lies in differentiating between solid and fluid cells in the discretized computational domain: 
A particle is mapped into the fluid simulation by marking the cells it occupies as solid, effectively excluding it from the LBM simulation.
This mapping has to be continuously updated for moving particles. 
This allows for a Lagrangian description of the particle, while the properties of the fluid use an Eulerian model.

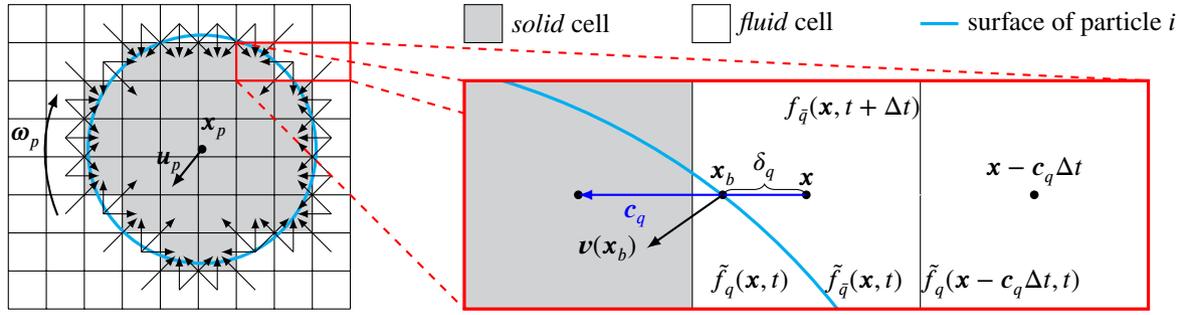
\begin{figure}[t]
	\centering
	\begin{tikzpicture}
	\coordinate[] (x1) at (2.55,2.1);
	
	\fill[lightgray] (1.5,1) rectangle ++(2,2.5);
	\fill[lightgray] (1,1.5) rectangle ++(0.5,1.5);
	\fill[lightgray] (2,0.5) rectangle ++(1,0.5);
	\fill[lightgray] (3.5,1) rectangle ++(0.5,2);
	
	\draw[step=0.5,black,thin] (0,0) grid (4.5,4);
	
	\draw[cyan,very thick] (x1) circle (1.5);
	
	\def\arrowLength{0.45}
	\foreach \point in {(1.75,0.75),(1.25,1.25),(0.75,1.75),(0.75,2.25),(0.75,2.75),(1.25,3.25)}
	{
		\draw[->,-latex] \point -- ++(\arrowLength,0.0);
	}
	\foreach \point in {(1.75,0.75),(1.25,1.25),(2.25,0.25),(2.75,0.25),(3.25,0.75),(3.75,0.75)}
	{
		\draw[->,-latex] \point -- ++(0,\arrowLength);
	}
	\foreach \point in {(3.25,0.75),(4.25,1.25),(4.25,1.75),(4.25,2.25),(4.25,2.75),(3.75,3.25)}
	{
		\draw[->,-latex] \point -- ++(-\arrowLength,0);
	}
	\foreach \point in {(1.25,3.25),(1.75,3.75),(2.25,3.75),(2.75,3.75),(3.25,3.75),(3.75,3.25)}
	{
		\draw[->,-latex] \point -- ++(0,-\arrowLength);
	}
	\foreach \point in {(3.25,0.75),(2.25,0.25),(1.75,0.75),(1.75,0.25),(1.25,1.25),(1.25,0.75),(0.75,1.25),(0.75,1.75),(0.75,2.25)}
	{
		\draw[->,-latex] \point -- ++(\arrowLength,\arrowLength);
	}
	\foreach \point in {(3.75,0.75),(3.25,0.75),(3.25,0.25),(2.75,0.25),(4.25,1.25),(4.25,1.75),(4.25,2.25),(4.25,0.75)}
	{
		\draw[->,-latex] \point -- ++(-\arrowLength,\arrowLength);
	}
	\foreach \point in {(2.25,3.75),(2.75,3.75),(3.25,3.75),(3.75,3.75),(3.75,3.25),(4.25,3.25),(4.25,2.75),(4.25,2.25),(4.25,1.75)}
	{
		\draw[->,-latex] \point -- ++(-\arrowLength,-\arrowLength);
	}
	\foreach \point in {(0.75,2.25),(0.75,2.75),(0.75,3.25),(1.25,3.25),(1.25,3.75),(1.75,3.75),(2.25,3.75),(2.75,3.75)}
	{
		\draw[->,-latex] \point -- ++(\arrowLength,-\arrowLength);
	}
	
	\draw[fill=black] (x1) circle (0.05) node[anchor=south,xshift=0.15cm] {$\boldsymbol{x}_p$};
	\draw[|->,-latex,thick] (x1) -- ++(-0.4,-0.5) node[pos=0.35,anchor=east] {$\boldsymbol{u}_p$};
	\draw[->,-latex,thick] (0.65,1.23) arc (202.5:157.5:2.1) node[anchor=east,yshift=-0.6cm,xshift=-0.1cm] {$\boldsymbol{\omega}_p$};

	\draw[red,thick] (3,3) rectangle ++(1.5,0.5);
	\draw[red,thick,dashed] (3,3) -- (6,0);
	\draw[red,thick,dashed] (3,3.5) -- (6,3);
	\draw[red,thick,dashed] (4.5,3) -- (15,0);
	\draw[red,thick,dashed] (4.5,3.5) -- (15,3);
	
	\fill[white] (6,0) rectangle ++(9,3);
	\fill[lightgray] (6,0) rectangle ++(3,3);
	\draw[step=3,black,thin] (6,0) grid ++(9,3);
	\draw[cyan,very thick] (6.5,3) arc (73:38.5:9);

	\coordinate[] (x) at (10.5,1.5);
	\coordinate[] (xMcq) at (13.5,1.5);
	\coordinate[] (xb) at (9.4,1.5);
	\coordinate[] (xs) at (7.5,1.5);
	\coordinate[] (ub) at (-1,-0.7);
	
	\draw[white] (x) -- ($(x)!0.5!(xs)$) node[pos=0.5,anchor=north,yshift=-0.8cm] {\color{black}$\tilde{f}_q(\boldsymbol{x},t)$};
	\draw[white] (x) -- ($(x)!0.5!(xMcq)$) node[pos=0.5,anchor=north,yshift=-0.8cm] {\color{black}$\tilde{f}_{\bar{q}}(\boldsymbol{x},t)$};
	\draw[white] (xMcq) -- ($(xMcq)!0.3!(x)$) node[pos=0.5,anchor=north,yshift=-0.8cm] {\color{black} $\tilde{f}_q(\boldsymbol{x}-\boldsymbol{c}_q\Delta t,t)$};
	\draw[white] (x) -- ($(x)!0.5!(xMcq)$) node[pos=0.4,anchor=south,yshift=0.8cm] {\color{black}$f_{\bar{q}}(\boldsymbol{x},t+\Delta t)$};
	
	\draw[fill=black] (x) circle (0.05) node[anchor=south] {$\boldsymbol{x}$};
	\draw[fill=black] (xMcq) circle (0.05) node[anchor=south] {$\boldsymbol{x}-\boldsymbol{c}_q \Delta t$};
	\draw[fill=black] (xb) circle (0.05) node[anchor=south] {$\boldsymbol{x}_b$};
	\draw[fill=black] (xs) circle (0.05);
	
	\draw[|->,-latex,blue,thick] (x) -- (xs) node[pos=0.75,anchor=north] {$\boldsymbol{c}_q$};
	\draw [decorate,decoration={brace,amplitude=4pt}] (xb) -- (x) node [pos=0.5,anchor=south,yshift=2pt] {$\delta_q$};
	
	\draw[->,-latex,thick] (xb) -- ($(xb)+(ub)$) node[pos=1,anchor=east] {$\bm{v}(\bm{x}_b)$};
	
	\draw[red,very thick] (6,0) rectangle ++(9,3); 
	
	\draw[fill=lightgray,thin] (6,3.5) rectangle ++(0.5,0.5);
	\node[right] at (6.5,3.75) {\textit{solid} cell};
	\draw[thin] (9,3.5) rectangle ++(0.5,0.5);
	\node[right] at (9.5,3.75) {\textit{fluid} cell};
	\draw[cyan,very thick] (12,3.75) -- ++(0.5,0.0);
	\node[right] at (12.5,3.75) {surface of particle $i$};
	
	\end{tikzpicture}
	\caption{Sketch of fluid-particle coupling, including the explicit mapping and the CLI boundary condition.}
	\label{fig:MappingAndCLIsketch}
\end{figure}

In this method, the hydrodynamic interaction force is evaluated as the force acting 
on the surface of the particle.
Within the particle only nodes marked as solid are present. 
Momentum is transferred solely between the 
particle boundary and the adjacent fluid, as sketched in the left part of~\cref{fig:MappingAndCLIsketch}. 
Along the fluid-particle boundary, the particle surface velocity 
acts as a no-slip boundary condition for the fluid.
We here employ the higher-order central linear interpolation (CLI) scheme~\cite{Ginzburg2008}, given by
\begin{equation}
	f_{\overline{q}}\left(\bm{x}, t + \Delta t \right) = \tilde{f}_q\left(\bm{x}, t\right) + \kappa_0 \tilde{f}_q(\bm{x}-\bm{c}_q\Delta t, t) - \kappa_0 \tilde{f}_{\overline{q}}\left(\bm{x}, t\right) - 3 \alpha \dfrac{w_q}{c_s^2} \rho_0 \bm{v}(\bm{x}_b, t) \bm{c}_q \text{,}
\end{equation}
with coefficients $\kappa_0 = \left(1-2\delta_q \right) / \left( 1 + 2 \delta_q\right)$, $\alpha = 4/\left(1+2\delta_q\right)$ and where $\overline{q}$ denotes the opposite direction of $q$. 
The variable $\delta_q$ describes the ratio between the distance from the fluid cell center to the boundary and the distance from the cell center to the solid cell center.
Therefore, the exact boundary location in direction of the lattice velocity is obtained as $\bm{x}_b = \bm{x} + \bm{c}_q \delta_q$.
This introduces subgrid information about the actual boundary location and substantially improves the accuracy to second order, as opposed to the commonly applied simple bounce-back rule that delivers only first order accuracy~\cite{Ginzburg2008,Rettinger2017}.
 
The link-based force contribution 
	$\bm{F}_{q_{f-s}}$
acting onto a particle at position 
	$\bm{x}_b$ via a fluid-solid link $q_{f-s}$
is given as~\cite{Ladd1994,Wen2014,Rettinger2017}
\begin{equation}
		\bm{F}_{q_{f-s}}(\bm{x}_b, t ) = \dfrac{(\Delta x)^3}{\Delta t} \left(\left( \bm{c}_{q_{f-s}} - \bm{v}(\bm{x}_b, t)\right) \tilde{f}_{q_{f-s}}\left(\bm{x}, t\right) - \left(\bm{c}_{\overline{q}_{f-s}} - \bm{v}(\bm{x}_b)\right)f_{\overline{q}_{f-s}}(\bm{x}, t+\Delta t) \right) \text{.}
\end{equation}

Following Ref.~\citenum{Wen2014}, the particle's surface velocity $\bm{v}(\bm{x}_b, t)$ is subtracted from the lattice velocities to ensure Galilean invariance. 
Summing up over all contributing nodes $q_{f-s}$, the total hydrodynamic force $\fhyd$ and torque $\thyd$ acting on one submerged particle are obtained:
\begin{equation}
	\fhyd = \sum_{\bm{x}_b} \sum_{q_{f-s}} \bm{F}_{q_{f-s}}(\bm{x}_b, t) \text{, }
	\thyd = \sum_{\bm{x}_b} \sum_{q_{f-s}} (\bm{x}_b - \bm{x}_p) \times \bm{F}_{q_{f-s}}(\bm{x}_b, t) \text{.}
\end{equation}

Updating the explicit particle mapping for a moving particle requires a careful reconstruction of fluid information in cells
that gets uncovered in this step.
We employ the approach presented in Ref.~\citenum{dorschner2015} which 
makes use of local velocity gradient information for an improved approximation.

\FloatBarrier

\section{Coupling techniques for light particles}
\label{sec:extension_light_particles}

\subsection{Limitations of Current Approach}
\label{sec:extension_original_shortcomings}

\begin{figure}
	\centering
\begin{subfigure}{0.45\textwidth}
	\centering
	\includegraphics[height=140pt]{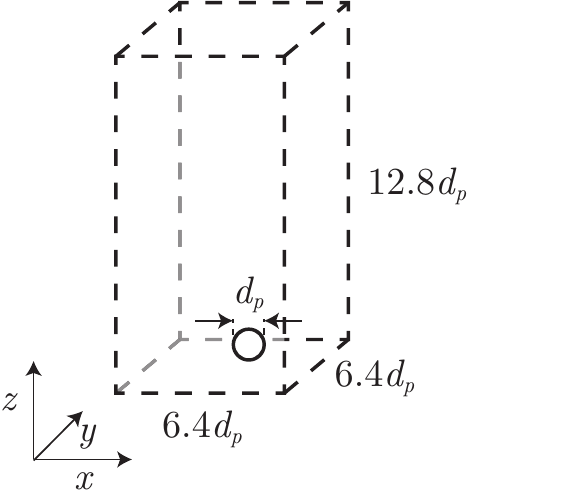}
	\caption{Layout of a simulation for a rising sphere in a periodic simulation domain. The sphere rises in $z$ direction.}
	\label{fig:introduction_linear_testcase_setup}
\end{subfigure}~
\begin{subfigure}{0.45\textwidth}
	\centering
	\includegraphics[height=140pt]{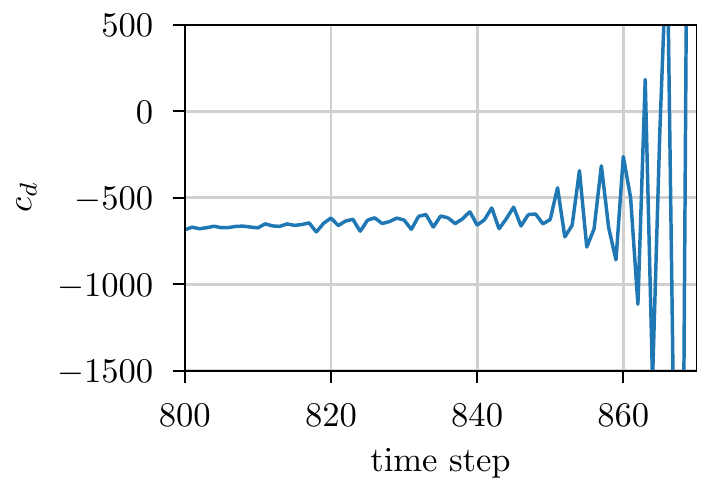}
	\caption{Oscillations in the drag coefficient $c_d$ at $\pip = 0.03$, $\Ga = 170$ and $d_p = 40$.} \label{fig:force_crashing_zoom}
\end{subfigure}
	\caption{Test case to study the stability of the coupling approach for a sphere rising inside a box.}
\end{figure}

To evaluate the approach outlined in Sec.~\ref{sec:numerical_methods} for
simulating a light sphere rising in a fluid, 
we consider the setup shown in \cref{fig:introduction_linear_testcase_setup},
with a density ratio $\pip = \rho_p / \rho_f < 1$.
The setup comprises a single submerged sphere of diameter 
$d_p$, initially residing at the bottom of the cuboidal 
domain with dimensions 
$(6.4, 6.4, 12.8)d_p$.
We employ periodic boundary conditions in all directions, and set  
the initial position of the particle 
to $(3.2, 3.2, 0.6)d_p$.
During the 
simulation, the particle rises 
in $z$ direction, as characterized by the Galilei number 
\begin{equation}
\Ga = \frac{u_g d_p} {\nu_f},\quad\text{ with } u_g = \sqrt{|\pip - 1| g d_p} \text{,}
\end{equation} 
where $g$ is the magnitude of the gravitational acceleration and $u_g$ is the characteristic, 
gravitational velocity.
In our simulations, we set $u_g = 0.01$ in lattice units and use $d_p$ to control the overall numerical resolution.
Depending on the values for $\pip$, $\Ga$, and the numerical resolution, 
the original method can predict the  
upward motion of the sphere or it may fail by becoming unstable.
In the latter case, we observe rapidly growing oscillations of the hydrodynamic force
acting on the sphere,
as illustrated in \cref{fig:force_crashing_zoom} for an exemplary case.
Here, the hydrodynamic force is expressed in terms of the dimensionless drag coefficient \cite{Holzer2008} $c_d = 2 \fhyd / \pip u_z \pi r_p^2$, where $u_z$ is the sphere's vertical velocity and $r_p = \nicefrac{1}{2} d_p$.
With increasing particle density this behavior is delayed to later stages of the simulation until completely disappearing at a certain density threshold. 
A simulation is considered stable, if no instabilities occurred until the rising sphere has reached its terminal ascension velocity at normalized time $t_n = t / t_g = 10$, with $t_g = d_p / u_g$.

\begin{table}[t]%
	\centering
	\caption{Lowest density ratios $\pi_{p,min}$ at $\Ga = 100$ and a certain sphere diameter $d_p$ for which stable results can be obtained with \mem-coupled LBM and particle phases.\label{tab:force_limit_results}}%
	\begin{tabular*}{300pt}{@{\extracolsep\fill}lccccccc@{\extracolsep\fill}}
		\toprule
		$d_p / \Delta x$			& 10    & 15    & 20   & 30   & 40  & 50 & 80 \\
		\midrule
		$\pi_{p,min}$	& 0.13  & 0.09  & 0.066 & 0.045  & 0.034  & 0.03 & 0.02 \\
		\bottomrule
	\end{tabular*}
\end{table}

\cref{tab:force_limit_results} lists the minimum density ratio $\pi_{p,min}$ for which we observe stable results at a certain $d_p$ for $\Ga = 100$.  
Most notably, $\pi_{p,min}$ behaves close to inversely proportional to the sphere diameter: 
For instance, at a diameter of 20, 0.066 is the lowest achievable value of $\pip$, while at the doubled diameter (40) approximately half the density ratio (0.034) is possible to simulate. 
The increase in diameter entails a finer resolution of the particle-fluid interface, such that stability issues occur at smaller density ratios.
However, every increase in resolution comes with a significant increase in computational costs, such that an alternative approach is required for simulating very light particles efficiently.
As such, these observations establish a baseline for possible improvements.

\subsection{Analysis of the Instabilities}

During each time step of our numerical method,
the hydrodynamic interaction forces and torques are first applied 
to the particle and then a time step for the particle is executed\cite{Rettinger2020b}.
Thus, these interactions 
always incorporate a time lag 
depending on the spatial and temporal resolution.
Additionally, also the liquid in the vicinity of a particle must be accelerated together with the particle.  
This is known as the \textit{added mass effect}. 
With an explicit fluid-particle coupling, however, this effect will be accounted for with a time delay. 
For heavy submerged particles, the added mass effect only plays a minor role and thus does not affect the dynamics. 
In contrast, for very light particles, the inertial forces originate mainly from the attached fluid mass being accelerated together with the particle~\cite{Schwarz2015}.
Hence, 
even though the inherent inaccuracies are minor, they may lead to oscillations 
whose amplitude 
grows.

This issue becomes clear by a closer inspection of the equations that describe the temporal evolution of  the linear and angular velocity of a single spherical particle submerged in fluid.
In general, these equations for geometrically fully resolving simulations are given as
\begin{equation}
\label{eq:trans_velocity_particle_equation}
\ddt{\bm{u}_p} = \dfrac{\rho_f}{m_p} \int_S \bm{\tau} \cdot \bm{n}\diff S + \dfrac{(\rho_p - \rho_f)V_p}{m_p}\bm{g} = \dfrac{1}{\pip V_p} \int_S \bm{\tau} \cdot \bm{n}\diff S + (\pip-1) \bm{g} = \dfrac{1}{\pip V_p} (\fhyd + \fg)
\end{equation}
and 
\begin{equation}
\label{eq:rot_velocity_particle_equation}
\ddt{\bm{\omega}_p} = \dfrac{\rho_f}{I_p} \int_S \bm{r} \times (\bm{\tau} \cdot \bm{n}) \diff S = \dfrac{1}{\nicefrac{2}{5} \pip V_p r_p^2} \int_S \bm{r} \times (\bm{\tau} \cdot \bm{n}) \diff S = \dfrac{1}{\nicefrac{2}{5} \pip V_p r_p^2} \thyd \text{.}
\end{equation}
Here, $\bm{\tau}$ is the hydrodynamic stress tensor, $\bm{n}$ the normal vector of the particle surface $S$, and $\bm{r}$ the vector from the particle center to a point on its surface $S$~\cite{Schwarz2015,Kempe2012}. 
For a spherical particle with radius $r_p$, $I_p$ is given as $\nicefrac{2}{5} m_p r_p^2$.

For very light particles, i.e. $\pip \rightarrow 0$, the coupling relations become singular due to a vanishing denominator in \cref{eq:trans_velocity_particle_equation,eq:rot_velocity_particle_equation}.
In other cases, a small denominator with $\pip \ll 1$ results in a large 
factor magnifying  
the fluid contribution. 
This amplifies 
inaccuracies originating from the numerical 
computation of $\fhyd$ and $\thyd$ in explicit fluid-particle coupling approaches~\cite{Schwarz2015}.

\subsection{Virtual Mass Correction for the LBM} \label{ch:vm_intro}

This shortcoming for simulations of light particles can be overcome by the \textit{virtual mass approach}
of Schwarz \textit{et al.} \cite{Schwarz2015}.
This approach 
has originally been developed 
for a classical 
Navier-Stokes-based solver in combination with the immersed boundary method (IBM)~\cite{Kempe2012}.
The therein applied IBM originally struggled with density ratios lower than $0.4$ due to oscillations in angular and translational velocities leading to a divergence within only few time steps \cite{Schwarz2015}.
This detrimental behavior is therefore similar to the one we observed for the \mem{}-coupled simulations.

Here, we aim to employ the same stabilization technique to the LBM with the MEM-based particle
coupling, following the arguments provided by Schwarz \textit{et al.}\cite{Schwarz2015} for the IBM.
Alike the IBM, our numerical method accounts for the same \cref{eq:trans_velocity_particle_equation,eq:rot_velocity_particle_equation}.
Following Schwarz \textit{et al.}\cite{Schwarz2015}, we introduce the concept of a virtual force 
\begin{equation}
	\fv = m_p^v \tfrac{\mathrm{d}\bm{u}_p}{\mathrm{d}t}.
\end{equation}
It uses the virtual mass $m_p^v = \cv \rho_f V_p$, which introduces the virtual mass coefficient $\cv>0$.
Adding the virtual force $\fv$ to both sides of \cref{eq:trans_velocity_particle_equation} and rearranging yields
\begin{equation}
	\label{eq:trans_velocity_particle_equation_virtual_force}
	\ddt{\bm{u}_p} = \dfrac{1}{(\cv + \pip) V_p} \int_S \bm{\tau} \cdot \bm{n}\diff S + \dfrac{\pip-1}{\cv + \pip} \bm{g} + \dfrac{\cv}{\cv + \pip} \ddt{\bm{u}_p}.
\end{equation}
A direct comparison with \cref{eq:trans_velocity_particle_equation_virtual_force} reveals, that this corresponds to a particle with altered mass $m_p + m_p^v$ onto which an additional force $\fv$ is acting that exactly compensates the effect of the increased mass. 
We explicitly note that this virtual force is different from the
added mass force, a physically observable effect 
which is already accounted for in the 
geometrically fully-resolved simulation 
model, as we employ it here.
The virtual mass coefficient $\cv$ may be 
determined on an empirical basis.
To attain a positive effect, $\cv$ has to be large enough depending on the density ratio. 
For spherical particles, Schwarz \textit{et al.} propose
to use $\cv = 0.5$ referring to the added mass coefficient of submerged spheres~\cite{Schwarz2015}.
In our case, values of 0.5 or 1 were found to be suitable, which also satisfy the condition of a lower limit recently reported by Tavanashad \textit{et al.} \cite{Tavanashad2020}.

In \cref{eq:trans_velocity_particle_equation_virtual_force} 
the derivative of the translational particle velocity appears on either side of the equation.
Physically, this is well justified, as the effect induced by the virtual mass should be exactly
suppressed by the counteracting force.
From an algorithmic perspective, this prevents an explicit temporal integration.  
As the acceleration in the current time step is to be computed and thus unknown,
the acceleration on the right hand side 
must be approximated to allow explicit integration.
Schwarz \textit{et al.} 
employ second order Lagrange interpolation
using the three most recent velocities in order to extrapolate the acceleration\cite{Schwarz2015}.
Here, we alternatively propose to use the particle's acceleration from the previous time step as computed during the Velocity Verlet procedure (\cref{sec:rigid_particle_motion}).
This essentially simplifies the scheme, reduces memory consumption, and showed the most promising results in our prestudies.
Denoting the previous translational acceleration as $\bm{a}_{p, t-1}$ then leads to 
\begin{equation}
	\ddt{\bm{u}_p} \approx \dfrac{1}{(\cv + \pip) V_p} \int_S \bm{\tau} \cdot \bm{n}\diff S + \dfrac{\pip-1}{\cv + \pip} \bm{g} + \dfrac{\cv}{\cv + \pip} \bm{a}_{p, t-1} \text{.}
	\label{eq:trans_velocity_particle_equation_virtual_force_w_acceleration}
\end{equation}

Similarly, the equation for angular motion \cref{eq:rot_velocity_particle_equation} is complemented with a virtual torque \begin{equation}
	\tv = \nicefrac{2}{5} \cvomega\rho_f V_p r_p^2 \tfrac{\mathrm{d}\bm{\omega}_p}{\mathrm{d}t},
\end{equation}
where the coefficient $\cvomega$ serves the same purpose as $\cv$ for the translational component. 
Substituting the angular acceleration on the right hand side for an estimate $\bm{\beta}_{p, t-1}$ and rearranging gives
\begin{equation}
	\ddt{\bm{\omega}_p} \approx \dfrac{1}{\nicefrac{2}{5} V_p r_p^2 (\cvomega + \pip)} \int_S \bm{r} \times (\bm{\tau} \cdot \bm{n}) \diff S + \dfrac{\cvomega}{\cvomega+\pip} \bm{\beta}_{p, t-1} \text{.}
	\label{eq:rot_velocity_particle_equation_virtual_force_w_acceleration}
\end{equation}
Consequently, \cref{eq:num_methods_particle_translational_basic,eq:num_methods_particle_angular_basic} now contain $\fv$ and $\tv$, such that $\fp = \fhyd + \fg + \fv$ and $\tp = \thyd + \tv$. The therein applied mass and moment of inertia are now given by $m_p = (\rho_p + \cv \rho_f)V_p$ and $I_p = \nicefrac{2}{5} (\rho_p + \cvomega \rho_f) V_p r_p^2$, respectively.

Finally, \cref{alg:vm_proc} outlines the complete simulation algorithm, featuring the steps introduced by the virtual mass correction in green. 
The update of the flow field via LBM, and the successive computation of the hydrodynamic interactions remains untouched, as well as the computation of the submerged weight force.
Then, the virtual force and torque are computed based on the translational and rotational accelerations of the previous time step. 
The integrator of the particle is supplied with the virtually increased mass to update the particle's position and velocities.
For the remainder of the paper, we will refer to this approach as \vmmem{}, in contrast to the original \mem{} from \cref{sec:numerical_methods} that does not apply the green parts.

\definecolor{vmhighlight}{rgb}{0.2,0.7,0.2} 
\begin{algorithm}
	\caption{Outline of \vmmem{} for a single submerged particle. \label{alg:vm_proc}}
	\begin{algorithmic}[1]
		
		\For{each simulation step}
		\State Perform LBM step.
		\State Compute hydrodynamic interactions, and average over two time steps.
		\State Set buoyant and gravitational force.
		\State \textcolor{vmhighlight}{Calculate virtual force and torque.}
		\State Update particle's translational and angular velocity \textcolor{vmhighlight}{with virtually increased mass.}
		\State Update the particle mapping into the fluid domain.
		\EndFor
		
	\end{algorithmic}
\end{algorithm}

\subsection{Validation}

For validating the translational and angular components of the virtual mass approach, two different scenarios are applied next.
For moderate density ratios the unchanged \mem{}, \cref{sec:numerical_methods}, is used as a reference, together with results from Ref.~\citenum{Schwarz2015} where similar validation setups were used. 

\subsubsection{Virtual Added Mass Force \label{sec:linear_validation}}

To validate \cref{eq:trans_velocity_particle_equation_virtual_force_w_acceleration},
we employ the same setup and parameterization as described in \cref{sec:extension_original_shortcomings}. 
A first set of simulations are carried out with Galileo number $\Ga = 100$ at 
density ratios $\pip \in \left\{ 0.1, 1.1 \right\}$ and a resolution of $d_p=40$, 
where the \mem{} serves as a reference. 
A final test run at $\pip = 0.001$, which is well below the minimal density ratio for which we expect stable simulations using \mem{} and a sphere diameter of 40 cells, employs a Galileo number of $170$ and reference data is extracted from Ref.\citenum{Schwarz2015}.
For \vmmem, we use $\cv = 1$ in all cases. 

The normalized ascension velocity of the particle is given as $u_{z,n} = u_{z} / u_g$. 
Its mean squared error $\overline{\epsilon}_{sq}$ and mean relative error $\overline{\epsilon}_{rel}$ with respect to a reference $\hat{u}_{z,n}$ evaluated at $m$ points in time are defined as
\begin{align}
\overline{\epsilon}_{sq} &= \tfrac{1}{m} \sum_i^m \left(u_{z,n,i} - \hat{u}_{z,n,i}\right)^2 \text{,} \\
\overline{\epsilon}_{rel} &= \tfrac{1}{m} \sum_i^m \dfrac{u_{z,n,i} - \hat{u}_{z,n,i}}{\hat{u}_{z,n,i}} \text{.}
\end{align}

Additionally, the relative error in the terminal velocity
is given as $\epsilon_{term,rel} = \left(u_{z,n,term} - \hat{u}_{z,n,term}\right) / \hat{u}_{z,n,term}$.

\begin{figure}[ht]
	\centering
	\begin{subfigure}[t]{0.5\textwidth}
	\includegraphics[width=\linewidth]{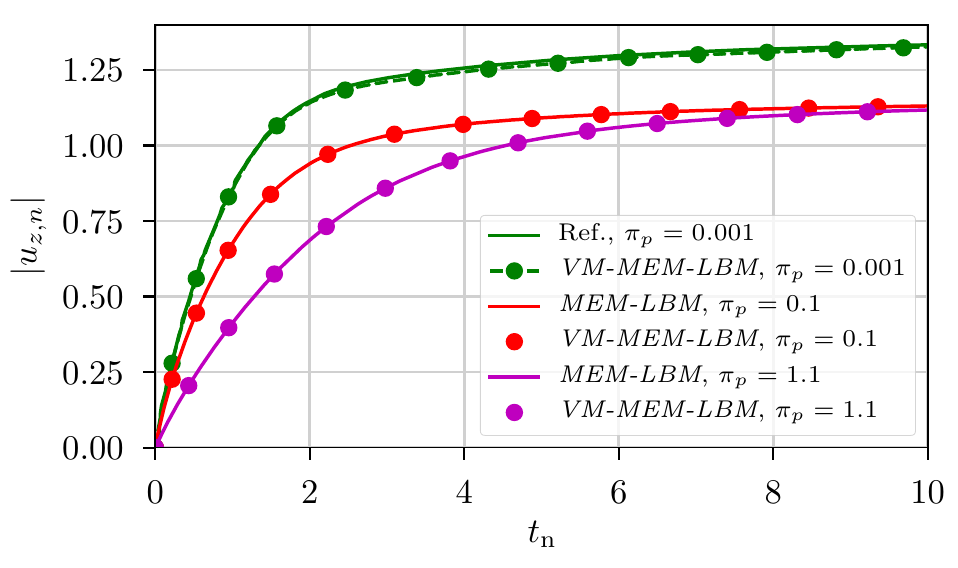}
	\caption{Complete run.}
\end{subfigure}%
	\begin{subfigure}[t]{0.5\textwidth}
	\centering
	\includegraphics[width=\linewidth]{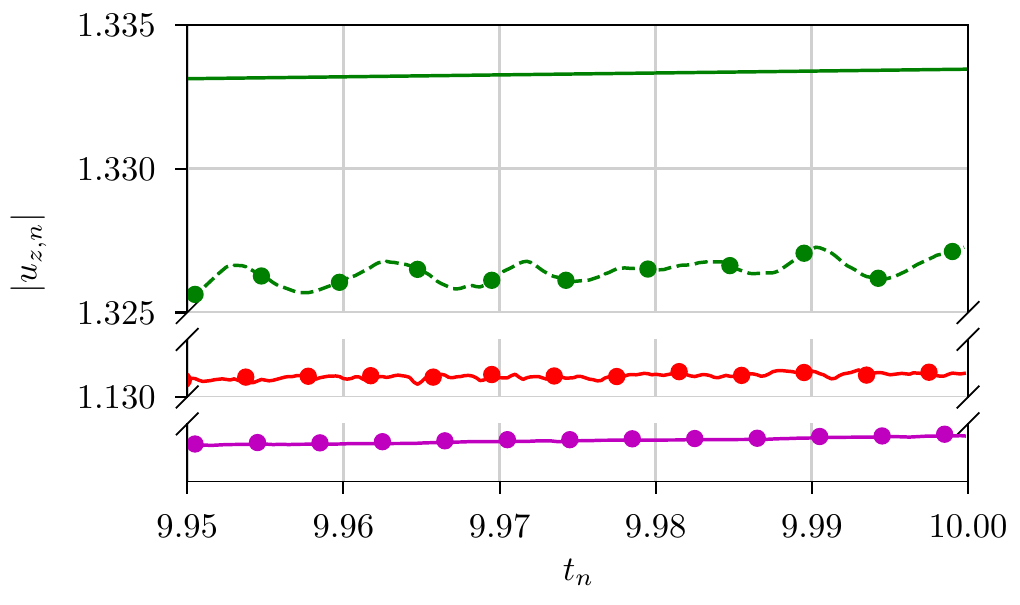}
	\caption{Magnification of end.}
\end{subfigure}
	\caption{Normalized rising velocities for the validation runs using \vmmem{} compared with their references. In case of $\pip = 0.001$, reference data is extracted from Ref.~\citenum{Schwarz2015}. For density ratios $> 0.001$, the \mem{} served as base line.}
	\label{fig:overall_linear_validaton_u_n}
\end{figure}

In \cref{fig:overall_linear_validaton_u_n}(a), the particle velocities of the simulations employing the \vmmem{} are compared with their reference. Overall, all density ratios show very good agreement during the whole simulation. Yet, $\pip = 0.001$ deviates slightly from the reference after $t_n \approx 2$.
The magnified view of the terminal velocities in \cref{fig:overall_linear_validaton_u_n}(b) shows light particles exhibiting an oscillating behavior with small amplitude of less than 0.1\% of $u_{z,n}$.
For heavier particles, these oscillations cannot be observed.

\begin{table}[t]%
	\centering
	\caption{Resulting particle Reynolds number $\Re_p = u_{z,term} d_p / \nu$ and errors in the normalized translational velocity $\vert u_{n} \vert$ when comparing the VM-MEM with reference data. Mean squared error $\overline{\epsilon}_{sq}$ and mean and terminal errors ($\overline{\epsilon}_{rel}$ and $\epsilon_{term,rel}$, respectively) relative to the reference velocities.\label{tab:linear_validation_overall_errors}}%
	\sisetup{
		exponent-product={\cdot},
	}
	\begin{tabular*}{340pt}{@{\extracolsep\fill}l
			S
			S
			S
			S
			@{\extracolsep\fill}}
		\toprule
		$\pip$ 						& \num{0.001} & \num{0.1}		& \num{1.1}		\\
		$\Ga$ 						& \num{170} & \num{100}		& \num{100}			\\
		\midrule
		$\Re_{p}$					& \num{225.64} & \num{113.08} & \num{111.77} \\
		\midrule
		$\overline{\epsilon}_{sq}$ 		& \num{0.01e-3} & \num{15.48e-9}	 	& \num{1.85e-9}	\\
		$\overline{\epsilon}_{rel}$			& \num{0.86e-3} & \num{0.28e-3}		& \num{0.15e-3}	\\
		$\epsilon_{term,rel}$		& \num{4.64e-3} & \num{0.04e-3}		& \num{0.06e-3}		\\
		\bottomrule
	\end{tabular*}
\end{table}

Visible in \cref{tab:linear_validation_overall_errors}, across all density ratios the \vmmem{} performs very well when compared with reference data. 
The mean relative errors $\overline{\epsilon}_{rel}$ amount to less than $0.3 \cdot 10^{-3}$ for all cases with a density ratio $\pip \geq 0.1$ (\cref{tab:linear_validation_overall_errors}). 
However, the mean squared error is much larger in the comparison of the simulation at $\pip = 0.1$. 
This results from fluctuations in the velocities, observable in \cref{fig:overall_linear_validaton_u_n}. 
In the terminal velocity, the relative error $\epsilon_{term,rel}$ for the three larger density ratios again is almost nonexistent, listed in \cref{tab:linear_validation_overall_errors}. 
At an increased terminal deviation of $\epsilon_{term,rel} = 4.64 \cdot 10^{-3}$, the error of the simulation with density ratio $\pip = 0.001$ is still within reasonable bounds.

Schwarz \etal{} initially report a relative error in the particle velocity of approximately $2 \%$ for their general scheme \cite{Schwarz2015}.
After adjusting the forcing point positions of the particle-fluid-coupling in their simulations they achieve a lower error of $2 \cdot 10^{-3}$, similar to our results.

\subsubsection{Virtual Added Mass Torque} \label{sec:angular_test_case}

\begin{figure}
	\centering
\begin{subfigure}[b]{0.5\textwidth}
	\centering
	\includegraphics[width=0.8\textwidth]{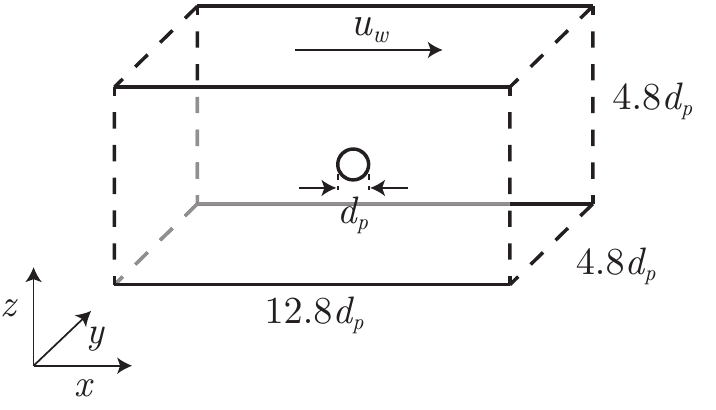}
	\caption{Domain setup.\\\hspace{\textwidth}}
	\label{fig:angular_testcase_setup}
\end{subfigure}~
\begin{subfigure}[b]{0.5\textwidth}
	\centering
	\includegraphics[width=1\textwidth]{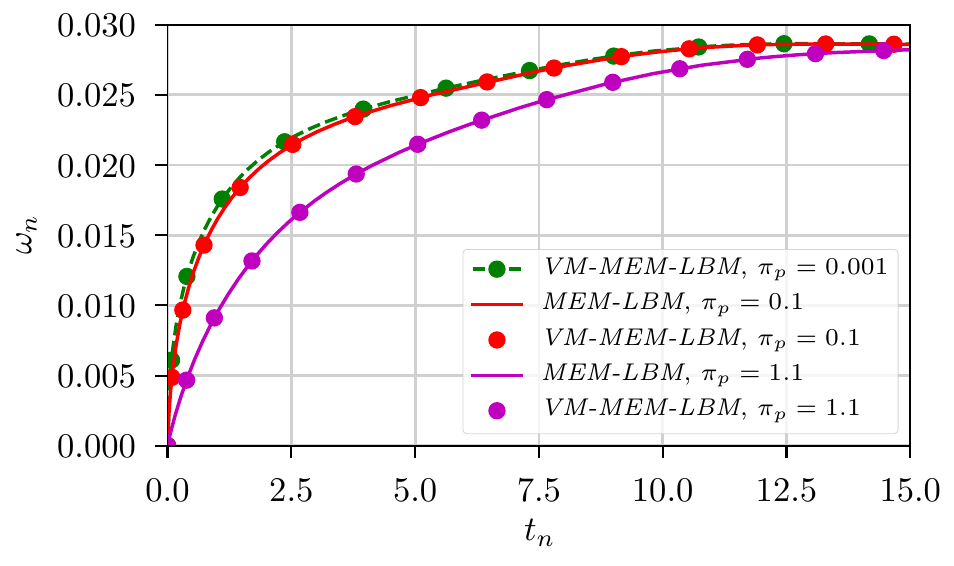}
	\caption{Normalized angular velocity $\omega_n$ for various density ratios, using either \mem{} or \vmmem{}.}
	\label{fig:overall_angular_validaton_u_omega_n}
\end{subfigure}
	\caption{Test case (\ref{fig:angular_testcase_setup}) and error (\ref{fig:overall_angular_validaton_u_omega_n}) to validate the \vmmem{} with respect to rotation. While the sphere's position is fixed throughout the simulation, the upper wall moving with velocity $u_w$ drives the Couette-initialized fluid, rotating the sphere after free rotation is unblocked.}
\end{figure}

Similar to Ref.~\citenum{Schwarz2015}, this test case features a domain of size $(12.8, 6.4, 4.8)d_p$, confined by two walls in $z$ direction and periodic otherwise. 
A sphere of $40$ cells in diameter is positioned in the domain center $(6.4, 3.2, 2.4)d_p$.
Here, the sphere is kept fixed at its initial position and all degrees of freedom are blocked at first. 
The fluid is initialized with a Couette profile, which is further driven by the upper moving wall at a constant velocity $u_w = 0.1$ in $x$ direction during the simulation.
This induces an angular momentum onto the particle, as sketched in \cref{fig:angular_testcase_setup}. 
After the flow field has fully developed, rotation of the particle is allowed and its rotational velocity over time is monitored.

The setup is defined by the density ratio $\pip$ and the particle Reynolds number $\Re_p = u_p d_p / \nu$,
where $u_p$ is the undisturbed fluid velocity at the sphere's position. 
Since the sphere resides in the center of the domain, the fluid velocity
evaluates to $u_p = \nicefrac{1}{2} u_{w}$.
The Reynolds number $\Re_p = 110$ is chosen to approximately 
match the 
situation in the force validation study in \cref{sec:linear_validation}. 
For $\pip \in \left\{ 0.1, 1.1 \right\}$, we again compare the results to \mem{}.
Additionally, $\pip=0.001$ is tested to check stability of \vmmem{} in contrast to the, for this case, unstable \mem{}.
We define $t_c = d_p / u_p$ and $\omega_c = u_p / d_p$
to normalize the time scale and the angular velocity, 
resulting in $t_n = t / t_c$ and $\omega_n = \omega_y / \omega_c$, respectively.

As
displayed in \cref{fig:overall_angular_validaton_u_omega_n}, 
simulations that employ the virtual mass correction agree very well to the reference data.
The mean relative errors
are below $0.5 \cdot 10^{-3}$ for both cases, see \cref{tab:angular_validation_overall_errors}. 
We point out that the terminal angular velocities match almost exactly. 
The proof-of-concept 
simulation at density ratio $0.001$ also appears to have performed well, 
with a 
plausible 
trend in angular velocity and, as expected, a stronger angular acceleration than the heavier spheres.

\begin{table}[t]%
	\centering
	\caption{Errors in the normalized angular velocity $\omega_n$ for \vmmem{} compared to simulations using \mem{}.\label{tab:angular_validation_overall_errors}}%
	\sisetup{
		exponent-product={\cdot},
	}
	\begin{tabular*}{200pt}{@{\extracolsep\fill}l
			S
			S
			S
			@{\extracolsep\fill}}
		\toprule
		$\pip$ 						& \num{0.1} & \num{1.1}		\\
		\midrule
		$\overline{\epsilon}_{rel}$			& \num{0.46e-3}		& \num{0.15e-3}		\\
		$\epsilon_{term,rel}$		& \num{0.01e-3}		& \num{0.0e-3}			\\
		\bottomrule
	\end{tabular*}
\end{table}

\subsection{Discussion}
We have demonstrated that the virtual mass  
corrections can be successfully adapted 
to the momentum exchange method.
Thus, the originally observed
oscillations in the motion of very light particles
are effectively reduced and stable simulations are obtained.
As the \vmmem{}
relies on an approximation of the particle's acceleration, the introduction of a certain error in comparison to \mem{} is unavoidable, as also observed in Ref.~\citenum{Schwarz2015}.
The validation scenarios reveal that this error is very small:
In the rising particle setup of \cref{sec:linear_validation}
the error stays below $0.5\%$, for the cases with $\pi_p \geq 0.1$ 
well below that. 
Evaluating the accuracy of angular motion in \cref{sec:angular_test_case},
the error thresholds for density ratios $\pi_p \in \{0.1, 1.1\}$ are similarly low at $0.05\%$, at most. 
Nevertheless, even though the virtual mass
was introduced as a solely numerically stabilizing technique for explicit fluid-particle coupling schemes,
the \vmmem{} might alter the physical results of the simulation. 
In phases with high acceleration changes, it artificially increases inertia by relying on the past acceleration value. 
Thus, for those phases, where the virtual force approximation might affect the accuracy negatively,
choosing a rather low value of $\cv$ is advisable.
In all our cases, values of $0.5$ or $1$ were found to be sufficient.
These arguments will be revisited in \cref{sec:numerical_study}, where the \vmmem{} again serves to simulate a sphere at $\pi_p = 0.001$ and results are compared to literature.

In a recent work, Tavanashad et al. \cite{Tavanashad2020} 
discussed the constant $\cv$ more extensively and established a lower positive limit for $\cv$, 
depending on the density ratio and the added mass coefficient. 
Our choice of $\cv$ is in line with this limit.

\section{Adaptive Grid Refinement}
\label{sec:agr}

Accurate numerical 
studies of 
particle rising in a fluid, as they will be carried out in \cref{sec:numerical_study}, 
require large computational domains to reduce the influence of the boundary conditions
on the trajectories.
On the other hand, ensuring an adequate representation of boundary layers
along the particle's surface and of the vortex structures
necessitates a fine numerical resolution in those regions.
Using a 
uniform grid for such simulations would result in enormous computational costs 
and limit the parameter space that can be explored substantially.
To alleviate this problem, we apply adaptive grid refinement in this work which significantly reduces the required computational resources.
In this section, we briefly outline and validate the 
grid refinement technique. 
The method makes use of the domain partitioning functionalities provided by the \walberla{} framework~\cite{schornbaum2016massively,Bauer2020}.

\subsection{Block-structured Grid Refinement with LBM}

Using a block-structured domain partitioning, the computational domain is split into so-called blocks, each containing a lattice of uniformly sized cells~\cite{Bauer2020}.
Grid refinement is then carried out by uniformly dividing a block into eight smaller blocks, if needed.
A block on the coarsest refinement level then has level $l = 0$, while blocks on the finest level have level $l = l_{max}$.
As each block features the same number of cells, the grid spacing $\Delta x_l$ 
of a block gets smaller with increasing level according to $\Delta x_{l+1} = \nicefrac{1}{2} \Delta x_l$.
Overall, a 2:1 balance between the blocks is maintained, meaning that the 
refinement level of neighboring blocks 
must at most 
differ by one.

A lattice Boltzmann method, capable of dealing with such non-uniform grids, was proposed by Rhode \textit{et al.} \cite{Rohde2006} and is applied here.
Details about its 
efficient implementation can be found in Refs.~\citenum{schornbaum2016massively,schornbaum2018extreme}.

\subsection{Refinement Criteria}

The refinement structure is defined by the individual target level of each block.
Criteria determining the target level are presented below. 
They are checked regularly throughout the simulation 
and their result is combined to satisfy all constistency requirements.
Subsequently, if some blocks 
are found to be on an unsuitable level of refinement, 
the grid is adapted by 
either coarsening or refining the blocks, obeying the 2:1 restriction.

\subsubsection{Particle-based Refinement Criteria}

\begin{figure}[t]
	\centering
	\includegraphics[width=0.8\textwidth]{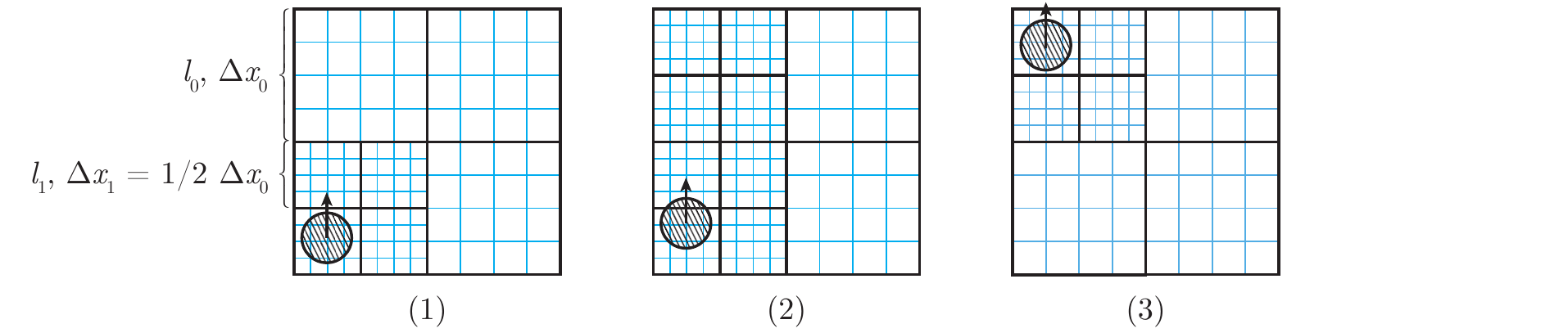}
	\caption{Refinement depending on the presence of a particle. Blocks are outlined black, containing 4x4 cells (outlined blue). (1) In the beginning, the block hosting the particle and its direct neighbors are on the finest level. (2) As soon as the particle enters the next block in vertical direction, the upper left non-refined block notices its presence in one of its lower neighbors and refines. (3) The particle now is fully contained within the upper quarter of the domain, such that the formerly four fine blocks at the bottom all are available to coarsen to one block. The frequency of the refinement evaluations has to be high enough to catch those particle movements, such that it always resides on the finest grid.}
	\label{fig:particle_refinement}
\end{figure}

During the simulation, we ensure that blocks that are in the vicinity of particles are always 
on the finest grid. 
This is motivated by both algorithmic and physical reasons.
From an algorithmic perspective, we 
avoid the 
software complexity of maintaining a consistent particle mapping 
across different refinement levels.
At the same time, this 
refinement rule  
ensures that the finest resolution is used to resolve  
the boundary layer around the particle. 
In these regions 
high velocity gradients occur and are thus resolved with maximal resolution.

Alternatively, in a block where no particle is nearby, 
the refinement level would be allowed to become coarser. 
Here other refinement criteria apply combined with the  2:1 balance between neighboring blocks.
A typical scenario is illustrated in \cref{fig:particle_refinement}.

\subsubsection{Flow Structure-based 
Refinement Criteria \label{sec:refinement_criteria}}

Secondly, 
special algorithmic \emph{sensors} are applied to the fluid to compute criteria for the refinement of a block. 
In our case, $\phi_c$, the scaled curl of the velocity field, is 
evaluated to detect shear layers and to 
control the resolution of 
high velocity gradients 
\cite{Crouse2003,Yuan2018,Deister1999}:
\begin{equation}
	\label{eq:sensor_curl}
	\phi_{c} = |\nabla \times \bm{u}_f| s ^{(r+1)/r} \text{.}
\end{equation}
Here, $\bm{u}_f$ denotes the cell local fluid velocity, $s$ a characteristic length scale for a cell
and $r$ a weighting coefficient. 
Multiplying by the weighted length scale $s^{(r+1)/r}$ facilitates the detection of weaker features in a coarser grid area, allowing weaker features to be refined when the stronger ones have been resolved \cite{Zeeuw1993}. 
The length scale $s$ can be chosen as $\sqrt[3]{V_{cell}}$ with $V_{cell}$ as the volume of a cell or, for cubical cells, simply $\Delta x$. 
The constant $r$ is commonly assigned a value of $2$ \cite{Crouse2003,Yuan2018,Deister1999}. 
A user-defined threshold $\sigma_c$ enables the adaptive manner of the fluid phase. 
Refinement is triggered when $\phi_c > \sigma_c$, while coarsening when $\phi_c < c\sigma_c$. 
To avoid repetitive refinement and coarsening, the value of $c$ should be lower than $1/2^{(r+1)/r}$, i.e. $c \lesssim 0.35$ for $r = 2$ \cite{Yuan2018}.

\subsection{Validation\label{sec:amr_tests}}

\begin{figure}[ht]
	\centering
	\begin{subfigure}[t]{0.3\textwidth}
		\centering
		\includegraphics[width=0.7\linewidth]{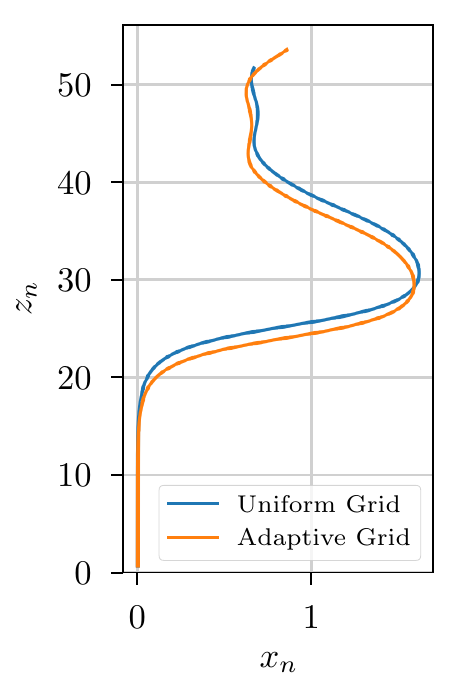}
		\caption{Particle displacement.}
		\label{fig:amr_comp_xz}
	\end{subfigure}~
	\begin{subfigure}[t]{0.3\textwidth}
		\centering
		\includegraphics[width=0.7\linewidth]{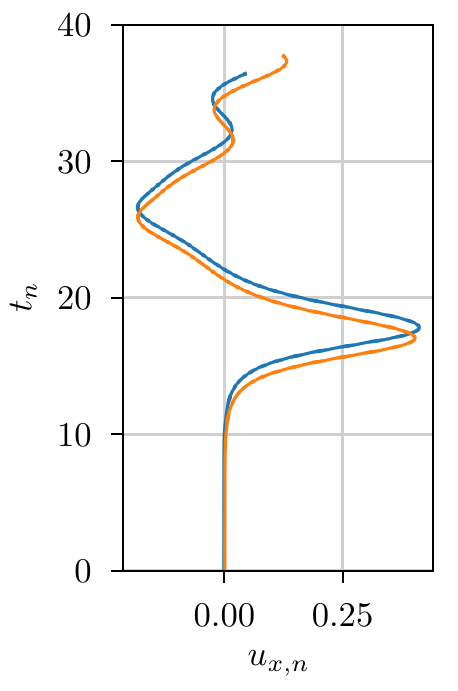}
		\caption{Velocity.}
		\label{fig:amr_comp_uxn}
	\end{subfigure}~
	\begin{subfigure}[t]{0.3\textwidth}
		\centering
		\includegraphics[width=0.7\linewidth]{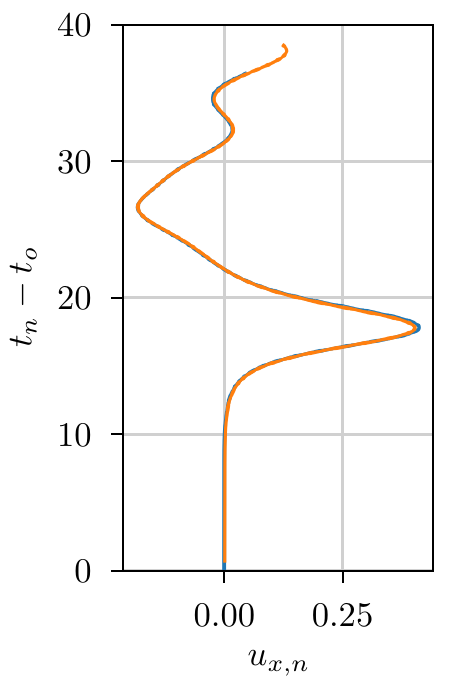}
		\caption{Aligned velocities.}
		\label{fig:amr_comp_uxn_aligned}
	\end{subfigure}
	\caption{Comparison of particle trajectory and velocity for adaptively refined (orange) and uniform (blue) simulation at $\Ga = 300$ and $\pip = 0.5$. Coordinates are normalized by particle diameter $d_p$, velocities by $u_g$.\label{fig:amr_comp}}
\end{figure}

To evaluate the accuracy and performance,
we 
compare adaptive grid refinement against a simulation using a uniform grid.
The 
parameters are chosen to reflect a typical setup 
as used in the next section.
In particular, we use the Galileo number $\Ga=300$ and a density ratio of
$0.5$ for a sphere of diameter $d_p=100$.
These properties do not require the application of \vmmem{} to stably couple solid and fluid phases, such that the \mem{} is employed.
The domain is 
fully periodic and of size (7.68, 7.68, 53.76)$d_p$. 
The sphere is positioned initially at (3.84, 3.84, 0.6)$d_p$.
Starting from this point, the sphere rises up to and reaches $z_n = z/d_p = 50$ 
at time $t_n = 36$. 
In that time span we can observe the beginning of a zig-zagging motion as depicted in \cref{fig:amr_comp}. 
Based on this scenario, we validate the sphere's trajectory in an adaptively refining grid 
against the one in a fully resolved reference simulation.

\begin{figure}[ht]
	\centering
	\begin{subfigure}[t]{0.48\textwidth}
		\centering
		\includegraphics[angle=-90,width=\textwidth]{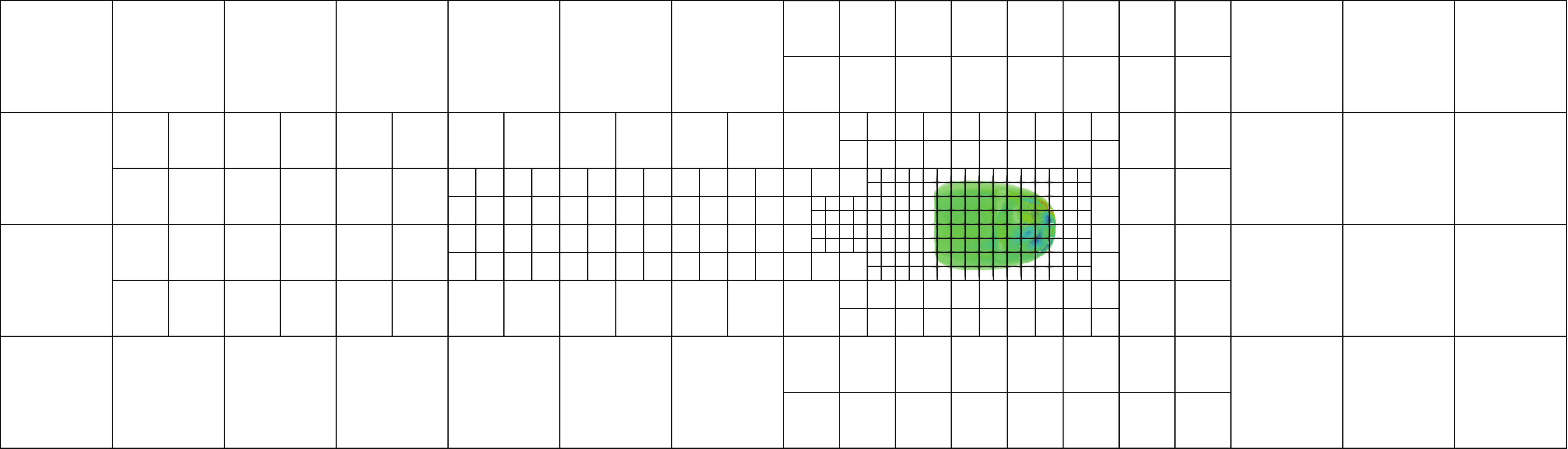}
		\caption{Adaptively refined grid.\label{fig:amr_domain_full_vs_adaptive_adaptive}}
	\end{subfigure}\hfill
	\begin{subfigure}[t]{0.48\textwidth}
		\centering
		\includegraphics[angle=-90,width=\textwidth]{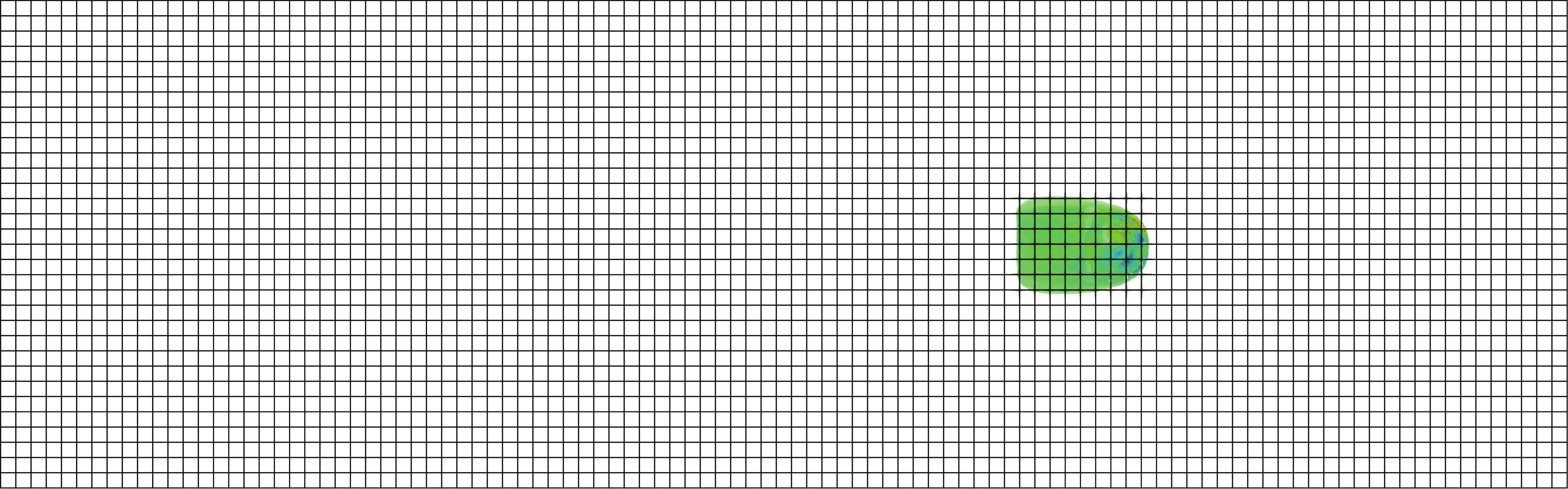}
		\caption{Uniform grid.\label{fig:amr_domain_full_vs_adaptive_full}}
	\end{subfigure}
	\caption{Domain partitioning at $t_n \approx 12.5$ in the adaptive and uniform case. The black lines depict the outline of the blocks. Only the lower region of the simulation domain is shown, where the primary movement direction ($z$) of the sphere is from left to right. The vortices around the particle are visualized using the Q criterion~\cite{Hunt1988}. \label{fig:amr_domain_full_vs_adaptive}}
\end{figure}

In the adaptively refined scenario, we set the refinement threshold $\sigma_c$ to $0.001$ and the coarsening factor $c$ to $0.2$. 
According to those values, a block is assigned one of 6 possible levels of refinement.
A visual comparison of the resulting domain partitioning for a small part of the domain is provided in \cref{fig:amr_domain_full_vs_adaptive_adaptive}, which shows the outline of the blocks, each containing $24^3$ cells.
The flow field around the rising sphere is visualized by the Q criterion~\cite{Hunt1988} for a time step right before the zig-zagging motion begins.

As the results show in \cref{fig:amr_comp}, the adaptively refined simulation behaves in the same manner as the one using a uniform grid after the initial trigger of the horizontal motion appears at a slight delay. 
The particle trajectory resides in a vertical plane, which is rotated by $45\degree$ in the horizontal plane.
\cref{fig:amr_comp_xz} shows the normalized vertical over the horizontal sphere displacement.

By defining a time $t_o$ at which the horizontal sphere velocity deviates from 0 and thus marks the beginning of the path-instability, we obtain a delay of $\Delta t_n = t_o^{uniform} - t_o^{refined} \approx 0.75$.
This observed delay is attributed to the unpredictable and to a certain degree random appearance of the path-instability. 
By accounting for this delay while plotting the velocity in \cref{fig:amr_comp_uxn_aligned}, the two lines essentially collapse into a single one.
A mean relative error between the aligned velocities of $1.1\%$ further underlines the agreement between both simulations and shows that the adaptive resolution introduces no substantial error in predicting the zig-zagging trajectory of the particle.

\begin{table}[ht]
	\caption{Comparison of setup and computational costs of the uniform and grid refinement scenario.}
	\label{tab:amr_domain_full_vs_adaptive_adaptive_setup}
	\centering
	\begin{tabular}{lccccc}
		\toprule
		Grid & (avg.) \#cells & (avg.) \#blocks & \#processes & run time (hours) & total core hours\\
		\midrule 
		Uniform & $3.06 \cdot 10^9$ & $221184$ & $36864$ & $10.27$ & $378474$ \\
		Adaptive & $3.78 \cdot 10^7$ & $2738.74$ & $768$ &  $6.58$ & $5337$ \\
		\bottomrule
	\end{tabular}
\end{table}

Regarding the computational resources of both simulations, the respective setup is described in Tab.~\ref{tab:amr_domain_full_vs_adaptive_adaptive_setup} and was run on the SuperMUC-NG supercomputer at LRZ in Garching.
Besides the reduction of the number of cells, the LBM for non-uniform grids carries out less time steps on coarser grids~\cite{Rohde2006} which further increases the efficiency of the grid refinement approach.
We find that the adaptively refined grid scenario reduced the required total core hours by a factor of 71.
In our case, we thus could obtain the essentially same results in less time and using significantly less processes.

As such, we have demonstrated that the here proposed grid refinement approach is capable of drastically reducing the computational costs of such a simulation while maintaining the accuracy of the uniform grid.
In combination with the virtual mass correction, this enables efficient simulations of rising particles with very small density ratios.

\section{Numerical study of rising particles}
\label{sec:numerical_study}

\begin{figure}
	\centering
	\begin{minipage}{0.50\linewidth}
		\centering
		\includegraphics[width=0.95\textwidth]{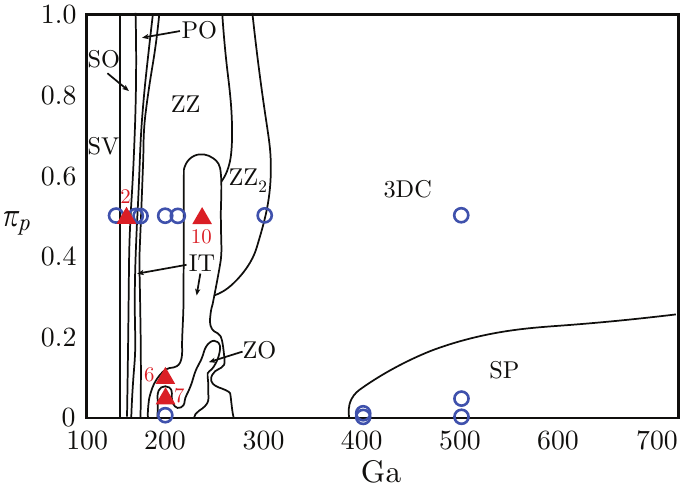}
	\end{minipage}%
	\begin{minipage}{0.50\linewidth}
    	\centering
		\resizebox{0.8\linewidth}{!}{
		\begin{tabular*}{0.9\linewidth}{c|l}
			\toprule
			& Description\\
			\midrule
			SV & Steady vertical\\
			SO & Steady oblique\\
			PO & Periodic oblique\\
			ZZ & Zig-zagging\\
			ZO & Oblique zig-zagging\\
			IT & Intermittent\\
			3DC & Three-dimensionally chaotic\\
			SP & Spiraling\\
			\bottomrule
		\end{tabular*}
		}
	\end{minipage}
	\caption{Regimes of particle motion characterized by density ratio $\pip$ and Galilei number $\Ga$ according to Ref.~\citenum{Auguste2018}. Overlaid points correspond to our simulations. Numbered triangles mark simulations at least partly differing from Auguste \& Magnaudet's results, circles largely confirm those.}
	\label{fig:auguste_ar_map_annotated}
\end{figure}

The trajectories of light ascending particles have attracted growing interest
since they may exhibit different regimes depending on certain parameters~\cite{Horowitz2010}.
With increasing computational power, also numerical simulations became possible~\cite{Jenny2003,Jenny2004,Biesheuvel2007,Zhou2015,Auguste2018}.
In their work from 2004, Jenny, Dušek \& Bouchet~\cite{Jenny2004}
showed that the type of trajectory is mainly determined by two parameters, the density ratio $\pip$ 
and the Galilei number $\Ga$, which describes the ratio of buoyant and viscous effects. 
They used numerical simulation to examine the particle paths that are encountered when varying 
$\pip \in [0, 10]$ and $\Ga \in [150, 350]$. 
In their work, they solve the Navier-Stokes equation based on a spectral-element
spatial discretization in a cylindrical domain moving with the sphere.
In particular, Jenny \etal{} isolated several regions in the parameter space, 
where rising spheres would show complex movements. 
The possible regimes range from steady vertical to three-dimensional chaotic,
with an oblique and oblique zig-zagging regime in between.

Horowitz \& Williamson~\cite{Horowitz2010} performed a 
comprehensive experimental study across a wide range of sphere densities and Reynolds numbers. 
Most notably, they identified a ``critical mass'' ratio, which marked the transition between a steady rise (for a sphere with a density ratio bigger than the critical value) and zig-zagging for a certain Reynolds number beyond the one corresponding to the loss of axisymmetry in the wake. 
The critical mass ratio amounted to $0.36 \pm 0.03$ for $\Re \in [260, 1550]$ and 
to $0.61$ for Reynolds numbers higher than $1550$. 

More recently, Auguste \& Magnaudet carried out an extensive set
of simulations using a spectral element method on a moving grid\cite{Auguste2018}
leading to results illustrated here in \cref{fig:auguste_ar_map_annotated}. 
Plotting the density ratio against the Galilei number, 
various patterns of motion emerge with increasing $\Ga$ and varying $\pip$.
Initially they encounter a steady vertical (SV) regime.
With further increasing Galilei numbers, steady oblique (SO) and periodic oblique (PO) motion follow. 
Planar zig-zagging (ZZ) arises next, where during each change of direction and between two crests vortices are shed. 
Depending on the density ratio, this regime reportedly persists up to $\Ga \approx 250$ 
for $\pip \gtrsim 0.6$ and only up to $\Ga \approx 220$ for $0.15 \lesssim \pip \lesssim 0.6$.  
Small amplitude zig-zags for spheres with density ratio of $\gtrsim 0.6$ 
in a three dimensional regime termed ZZ$_2$ follow next; 
at lower density ratios an intermittent regime (IT) up to $\Ga \approx 250$ occurs first. 
In a small stride of the map, very light spheres perform oblique zig-zagging (ZO). 
After a threshold value of $\Ga \approx 250$ to $300$, 
three dimensional chaotic (3DC) paths are assumed by submerged spheres of all density ratios. 
Beyond $\Ga \approx 400$ however, 
some regularity is regained in the spiraling motion (SP) for very light spheres. 

Except for some differences at $\Ga=200$ to $250$, 
these findings agree largely with the ones from Zhou \& Dušek 
\cite{Zhou2015}, 
who used a similar numerical method.
On the other hand, these numerical results show deviations from
the experimental studies of Horowitz \& Williamson, who e.g. do not record sphere trajectories pertaining to the 3DC regime.

In the present work, we study 16 combinations of 
$\Ga$ and $\pip$ that we simulate with the new method as proposed in this article, if required.
This allows us to further validate our method on the one hand, and to investigate the parameter spaces where existing studies report mismatching results.
The analyzed parameter combinations are marked
in \cref{fig:auguste_ar_map_annotated}. 

\subsection{Simulation Setup}

All simulations 
are performed in a cuboidal 
simulation domain with periodic boundary conditions in all directions.
Large enough domains are chosen
to avoid that turbulent structures spread so far that they would influence the particle wake 
or the trajectory by re-entering on the opposite side of the domain. 
Depending on the combination of $\Ga$ and $\pip$,
the sphere is resolved using a varying number of cells per 
diameter, expressed by $d_p/\Delta x$, to assure a fine enough resolution of the boundary layer along the sphere surface and the wake~\cite{Rettinger2017}.
We adjust the parameter $\Lambda_b$ of the MRT collision operator as introduced in 
\cref{sec:lbm} to $\Lambda_b = 1$, 
which recovers the TRT collision operator.

\begin{table}%
	\caption{Domain properties and sphere resolution $d_p / \Delta x$ used for the simulations of this chapter. Refinement check frequency is expressed in terms of time steps on the coarsest grid level. \label{tab:sim_setups}}%
	\begin{center}
		\begin{tabular*}{370pt}{lcccccc}
			\toprule
			Setup & Domain Size & $d_p / \Delta x$ & Refinement Levels & Refinement Check Freq.\\
			\midrule
			I	& (38.4, 38.4, 192) & 80 & 8 & 4\\
			I$_l$	& (38.4, 38.4, 384) & 80 & 8 & 4\\
			II	& (30.72, 30.72, 215.04) & 100 & 8 & 4\\
			III	& (38.4, 38.4, 307.2) & 160 & 9 & 8\\
			\bottomrule
		\end{tabular*}	
	\end{center}
\end{table}

In all cases, adaptive grid refinement, as described in \cref{sec:agr}, is applied to reduce the computational cost.  
Depending on the domain size, 8 or 9 levels of refinement are applied.
The necessity for an adaption of the grid is evaluated every 4 to 8 time steps on the coarsest grid level,.
The flow-based refinement criterion from \cref{sec:refinement_criteria} uses $\sigma_c = 0.001$ and $c = 0.2$.
Similar to the simulations described in \cref{sec:extension_light_particles}, 
the gravitational velocity $u_g = 0.01$ in lattice units is used 
which then determines the gravitational acceleration and the kinematic viscosity for the simulation.
A summary of the domain size, $d_p$ and refinement levels for each 
of the setups is presented in \cref{tab:sim_setups}.
The virtual mass approach from \cref{sec:extension_light_particles} in the fluid-solid coupling is only employed when it is required for stable simulations. 
In the present cases, this is in settings where the particle density $\pip = 0.001$.
Then, the virtual mass coefficients are chosen as $\cv = \cvomega = 0.5$.

\subsection{Simulation Results\label{sec:simulation_results}}

The 16 simulations 
range over of Galilei numbers 
from $150$ to $500$ at particle densities of $0.5$,  
$0.1$, $0.05$, $0.01$ and $0.001$. 
The results for the key indicators 
are concisely summarized in \cref{tab:results_rising_sphere}. 
The parameters are selected such that the 
range of regimes 
from the existing literature is covered.
In particular, we follow \AM's extensive study
and perform a subset of their simulations, plus two additional cases (Case 13 and 16 in \cref{tab:results_rising_sphere}).

Cases 1, 3 and 4 match with the descriptions by \AM~well, 
pertaining to the simpler patterns of steady vertical or oblique movement. 
However, a disagreement is found for Case 2
where 
the Galilei number is specifically chosen to be located 
in the transition regime between SV and SO.
\AM~identify the SO regime here,
while 
in our simulation the sphere appears to move SV, with visible transitions to an oblique motion.

In the next four cases, 5 to 8, we fix 
the Galilei number at 200, merely varying the density ratio. 
Here we observe three different
regimes: ZZ, 3DC, and PO.
While we find agreement 
with \AM~for Cases 5 and 8, 
simulations 6 and 7 do not
produce the PO trajectories as reported by \AM.
Instead, we observe a chaotic motion of the sphere, only in Case 6 initially resembling a PO path.
However, \ZD~also categorize those into the 3DC regime.

Further increasing the Galilei number and setting 
the density ratio to $0.5$ in Cases 9 to 11 shows 
agreement with \AM.
In addition to the already established ZZ regime 
in Cases 9 and 10, a second zig-zagging motion, ZZ$_2$, is identified in Case 11.
This is characterized by a higher frequency and a smaller amplitude.

The final five scenarios feature the largest Galilei numbers.
They all exhibit a spiraling motion, except Case 14.
There, the trajectory loses all regularity resulting in three-dimensional chaotic motion.
These findings are again in agreement with \AM. 
Additionally, \ZD~confirms our 
results in Cases 13 and 16, since our results for very light spheres (density ratio of $0.001$) match their results for massless spheres.

All in all, we identify similar issues as both previous numerical studies in replicating the experiments of Horowitz \& Williamson. 
A ``critical mass ratio'' and zig-zagging motion instead of three dimensional chaotic motion at increasing Reynolds numbers could not be found.

\begin{table}
	\caption{
	Results for simulations of a rising sphere at various combinations of Galilei number $\Ga$ and particle density ratio $\pip$. For comparison, regimes as found by \AM~(AM) and \ZD~(ZD) are provided next to ours. Abbreviations are explained in \cref{fig:auguste_ar_map_annotated}. Columns labeled $u_{z,t}$, $t_{n,t}$ and $z_{n,t}$ list the normalized terminal values of the respective quantity. $d_{cr.}$ and $h_{cr.}$ describe the lateral distance and height from one crest to the following crest after completing one iteration of either a spiraling or a zig-zagging path. Regimes from literature are mostly backed by simulations in the respective case; instances, where only a visual classification by diagram is available, are marked with (*).
	\label{tab:results_rising_sphere}}
\resizebox{\linewidth}{!}{
\begin{tabular}{cc|cc|ccccccccccc}
\toprule
\multirow{2}{*}{Case} & \multirow{2}{*}{Setup} & \multirow{2}{*}{Ga} & \multirow{2}{*}{$\pip$} & \multicolumn{3}{c}{Regime} & \multirow{2}{*}{Re} & \multirow{2}{*}{Incl. [$\degree$]} & \multirow{2}{*}{St} & \multirow{2}{*}{$d_{cr.}$} & \multirow{2}{*}{$h_{cr.}$} & \multirow{2}{*}{$u_{z,t}$} & \multirow{2}{*}{$z_{n,t}$} & \multirow{2}{*}{$t_{n,t}$} \\
                      &                        &                     &                         & Present     & AM     & ZD     &                     &                                &                     &                            &                            &                            &                            &                            \\
\midrule
1&I&150&0.5&SV&SV&SV&196.2&&&&&1.309&191.8&147.9\\
2&I&162.5&0.5&SV&SO&SO&218.2&&&&&1.343&191.8&143.1\\
3&I&172&0.5&PO&PO&SO&232.0&4.62&0.043&&&1.348&191.8&144.3\\
4&I$_l$&175&0.5&PO&ZZ,PO&PO&237.0&4.74&0.042&&&1.350&354.5&262.6\\
5&II&200&0.5&ZZ&ZZ&ZZ&277.8&&0.018&1.46&&1.408&214.8&155.2\\
6&II&200&0.1&3DC&PO&3DC*&277.5&&&&&1.356&214.8&155.1\\
7&II&200&0.05&PO, 3DC&PO&3DC*&277.9&2.90&&&&1.398&214.8&155.3\\
8&II&200&0.001&PO&PO&3DC&276.9&3.07&0.045&&&1.405&214.8&155.6\\
9&II&212.5&0.5&ZZ&ZZ&3DC&299.1&&0.028&1.14&&1.421&241.5&172.6\\
10&II&237.5&0.5&IT, ZZ&ZZ&3DC&344.5&1.34&0.098&0.22&&1.449&254.9&177.9\\
11&II&300&0.5&ZZ2&ZZ2&3DC&447.9&&0.099&0.28&&1.495&238.3&162.1\\
12&III&400&0.01&SP&SP&3DC/SP*&591.0&&0.070&0.98&14.20&1.475&171.9&117.3\\
13&III&400&0.001&SP&SP*&SP&586.6&&0.070&1.02&14.06&1.466&294.9&201.0\\
14&III&500&0.5&3DC&3DC&3DC&815.2&&&&&1.600&167.2&105.0\\
15&III&500&0.05&SP&SP&3DC/SP*&757.3&&0.067&1.08&13.81&1.510&266.3&177.6\\
16&III&500&0.001&SP&SP*&SP&729.3&0.20&0.074&1.09&13.22&1.455&264.8&181.0\\
\bottomrule
\end{tabular}
}
\end{table}

\subsection{Detailed Analysis of Results}

In the following subsections, we describe our simulation results in more detail, following the analysis provided by \ZD{} and \AM.
This enables an in-depth comparison with these studies, based on various quantities of interest that are also provided in \cref{tab:results_rising_sphere}.

In the following, all reported particle displacements are normalized in terms of the sphere diameter $d_p$. 
As such, normalized coordinate axes corresponding to the particle's trajectory are denoted as $x_n$, $y_n$ and $z_n$.
To quantify periodic paths, a dominant frequency $f$ can be defined and expressed as the dimensionless Strouhal
number $\St = f d_p / u_{z,t}$, 
see Ref. \citenum{Auguste2018}. 
Here, the quasi-steady vertical velocity $u_{z,t}$ during the rise of the particle serves as 
reference velocity, normalized beforehand using the gravitational velocity $u_g$. 
The frequency of sphere oscillations is  
computed by a Fast Fourier transformation 
of the horizontal component of the velocity 
and
extracting the frequency with the highest amplitude.
For comparison with \ZD, we 
compute the normalized frequency as $f_n = f t_g$.
Further, the lateral displacement $d_{cr.}$ and height $h_{cr.}$ are determined from one typical ``zig-zag'' or one circle of the spiral.
Inclination angles with respect to the vertical axis are computed by fitting a straight line by means of linear regression, starting at the beginning of the oblique displacement.

\subsubsection{Galilei Numbers $\Ga \leq 175$: Steady Vertical and Oblique}

\begin{figure}
	\captionsetup[subfigure]{position=b}
	\centering
	\subcaptionbox{Steady vertical and oblique ascent. \label{fig:rising_XZ_all}}{\includegraphics[width=.3\linewidth]{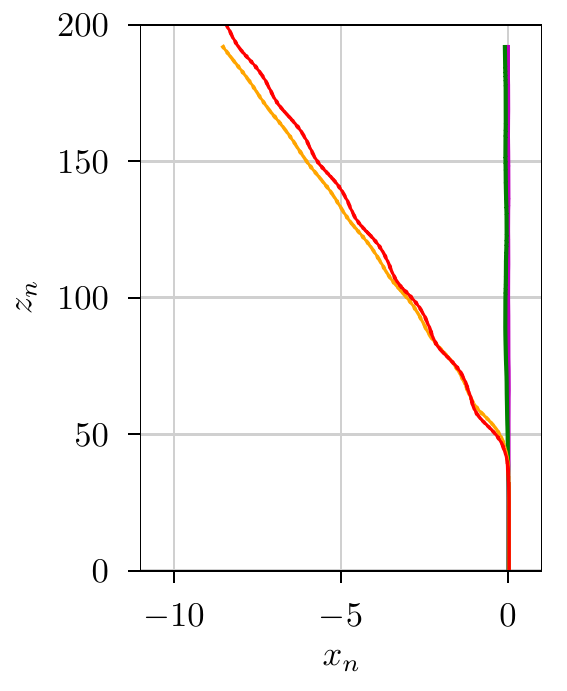}}
	\hfill
	\subcaptionbox{Minor lateral movement for Cases 1 and 2 ($\Ga = 150$ and $162.5$). \label{fig:rising_XYZ_150_162.5}}{\includegraphics[width=.3\linewidth]{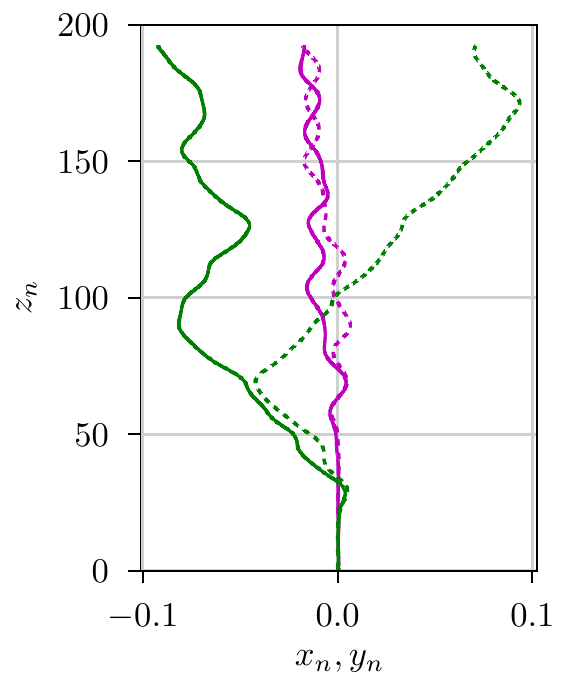}}
	\hfill
	\subcaptionbox{Wake structure for Case 3. At borders of differently refined regions visualization artifacts formed.\label{fig:ga_172_wake}}{\includegraphics[width=.35\linewidth]{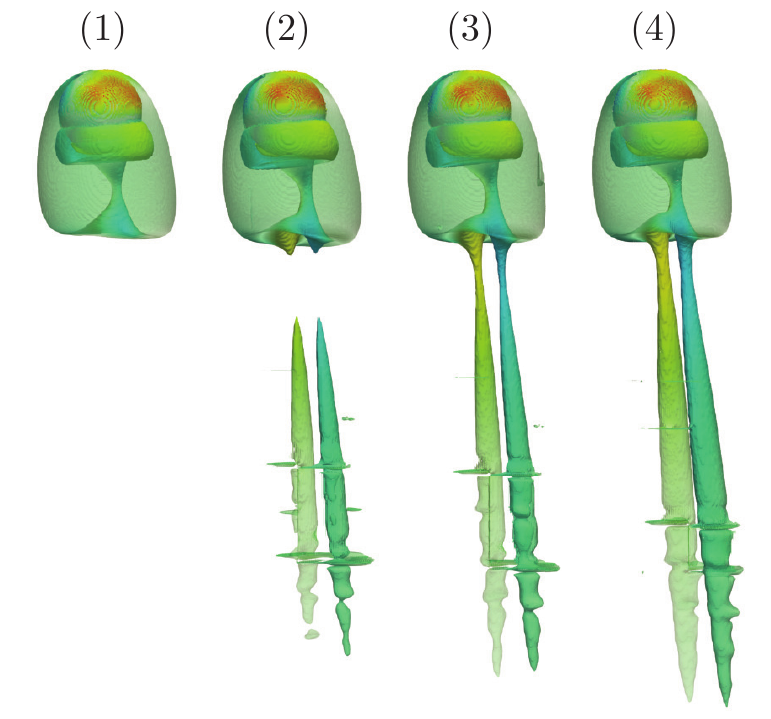}}
	\caption{Cases 1 to 4: Plots \textbf{(a, b)} show trajectories at $\pip = 0.5$: $\Ga = 150$ (magenta), $162.5$ (green),  $172$ (orange) and $175$ (red). Solid lines correspond to movement in $x$ direction, dashed in $y$ direction. In \textbf{(c)} the wake structure of Case 3 is visualized using the Q criterion at a threshold of $10^{-10}$ and colored by normalized vorticity (between $-3.3$, blue, 0 in green and $3.8$, red).}
\end{figure}

Cases 1 to 4 of \cref{tab:results_rising_sphere} feature Setup I, covering Galilei numbers from 150 to 175 at a density ratio of $0.5$.
We observe steady vertical and oblique movement for these simulations.

At Galilei number 150 (Case 1), the sphere rises in a steady vertical line, with only very minor lateral displacement. 
This is in line with findings of both \AM~and \ZD.

Case 2 at $\Ga = 162.5$ still shows movement corresponding to the SV regime.
Yet for \AM, this case marked the transition to a 
second regime with steady oblique movement. 
A closer inspection of our obtained trajectory hints at a transition to obliqueness (\cref{fig:rising_XYZ_150_162.5}),
which supposedly is triggered at a slightly higher Galilei number. 
Unlike Case 1, the path is less oscillating around a vertical axis,
but shows especially for the $y$ direction obliqueness with lateral displacement in the order of one tenth of the sphere diameter.

Oblique paths are first observed for Case 3 at $\Ga = 172$. 
This is the first occurrence of a PO path, and corresponds well with
\AM.
Further considering the inclination angle between the trajectory and the vertical axis, 
\AM~reported an average 
value of approximately $4.5\degree$, in comparison to $4.6\degree$ in our case.
The comparison Strouhal number $\St = 0.05$ is close to the one at hand ($0.043$). 
\cref{fig:ga_172_wake} depicts the wake behind the rising particle:
The sphere is constantly 
enclosed by a flow that builds up and sheds two axisymmetric fluid tubes in regular intervals.

The last case, 4, uses the enlarged Setup I$_l$ to capture a longer impression of the trajectory. 
The general motion stays the same compared to Case 3,
with the inclination slightly changing to $4.74 \degree$. 
\AM here noticed zig-zagging at first, which, however, ends
after some iterations and changes to an oblique movement, like observed here right from the beginning.

\subsubsection{Galilei Numbers $200 \leq \Ga \leq 300$: (Oblique) Zig-Zagging and Three Dimensionally Chaotic}

\begin{figure}
	\captionsetup[subfigure]{position=b}
	\centering
	\hfill
	\subcaptionbox{3D plot of trajectories at $200 \leq \Ga \leq 300$.\label{fig:rising_sphere_180_ga_300_all_xyz}}
	{\includegraphics[width=.35\linewidth]{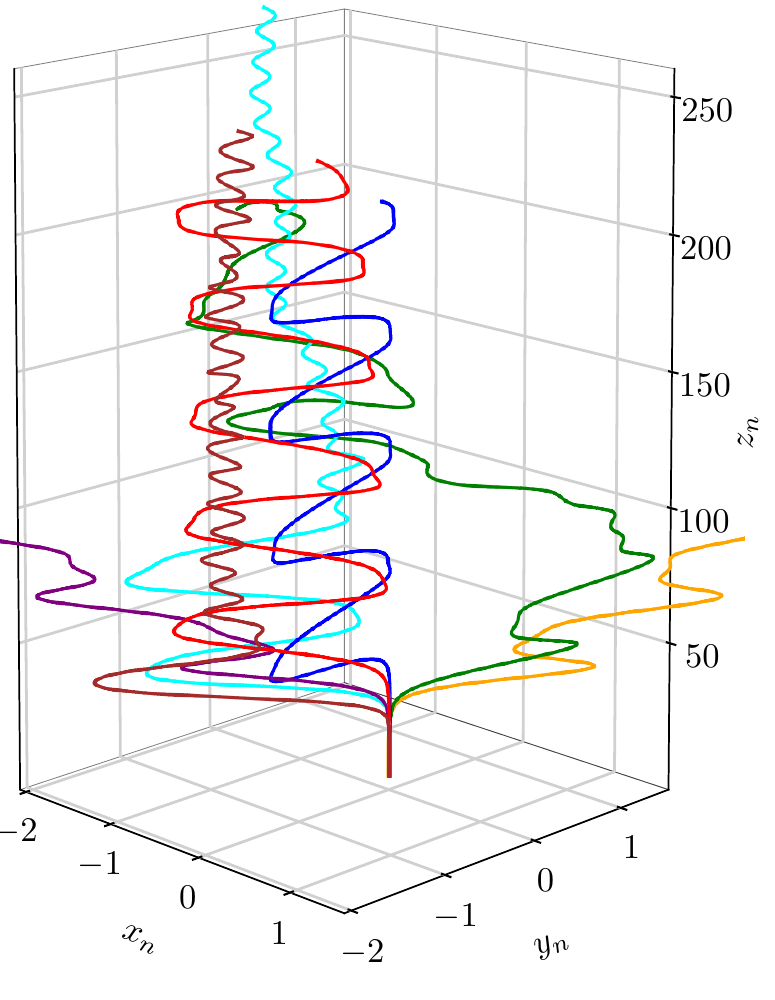}}
	\hfill
	\subcaptionbox{$\Ga = 200$, $\pip = 0.001$ (purple) and $\pip = 0.05$ (orange, mirrored at $(0,0)$).\label{fig:rising_sphere_ga_200_pi_0.05_and_0.001_xyz}}{\includegraphics[width=.35\linewidth]{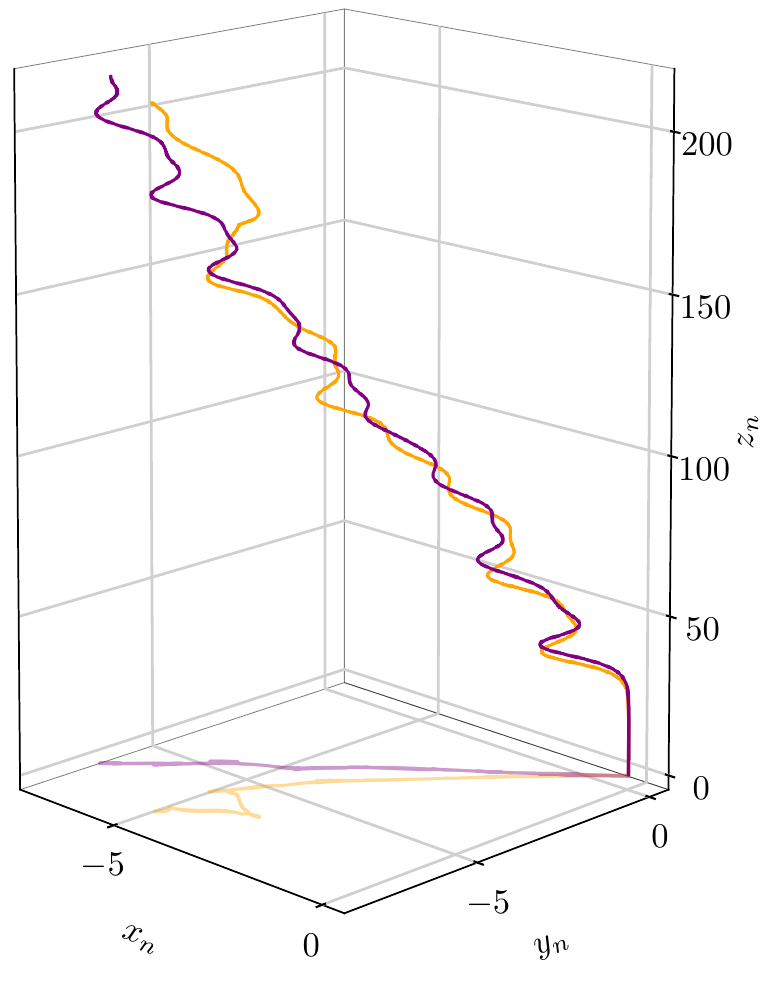}}
	\hfill
	\hfill
	\caption{Cases 5 to 11: Trajectories for the range of Galilei numbers between 200 and 300. Combinations of Galilei number and density ratio depicted as follows. $\Ga = 200$: $\pip = 0.5$ (blue), $\pip = 0.1$ (green), $\pip = 0.05$ (orange), $\pip = 0.001$ (purple); $\Ga = 212.5$: $\pip = 0.5$ (red); $\Ga = 237.5$: $\pip = 0.5$ (cyan); $\Ga = 300$: $\pip = 0.5$ (brown).
	}
\end{figure}

A series of six simulations (Cases 5 to 11) using Setup II 
features Galilei numbers between 200 and 300 at various density ratios, as low as $\pip = 0.001$.
Following these parametric alterations, vertical and oblique zig-zagging and three dimensionally chaotic regimes emerge.

\begin{figure}
	\captionsetup[subfigure]{position=b}
	\centering
	\hfill
	\subcaptionbox{Case 5: $\Ga = 200, \pip = 0.5$: ZZ.\label{fig:rising_sphere_ga_200_pi-0.5_xy}}[0.3\linewidth]{\includegraphics[width=0.8\linewidth]{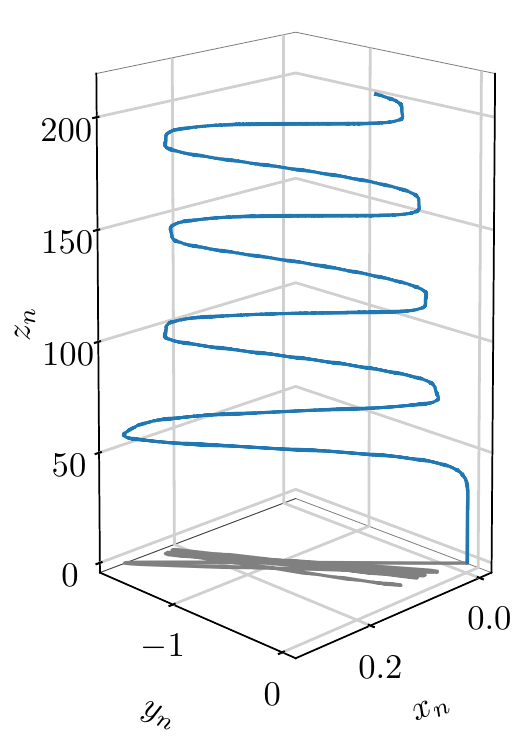}}
	\hfill
	\subcaptionbox{Case 10: $\Ga = 237.5, \pip = 0.5$: ZZ and ZZ$_2$.\label{fig:rising_sphere_ga_237.5_pi-0.5_xy}}[0.3\linewidth]{\includegraphics[width=0.8\linewidth]{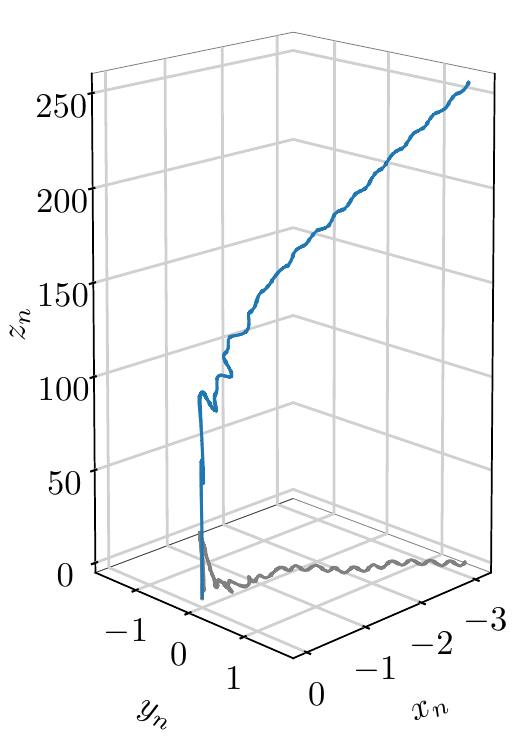}}
	\hfill
	\subcaptionbox{Case 11: $\Ga = 300, \pip = 0.5$: ZZ$_2$.\label{fig:rising_sphere_ga_300_pi-0.5_xy}}[0.3\linewidth]{\includegraphics[width=0.8\linewidth]{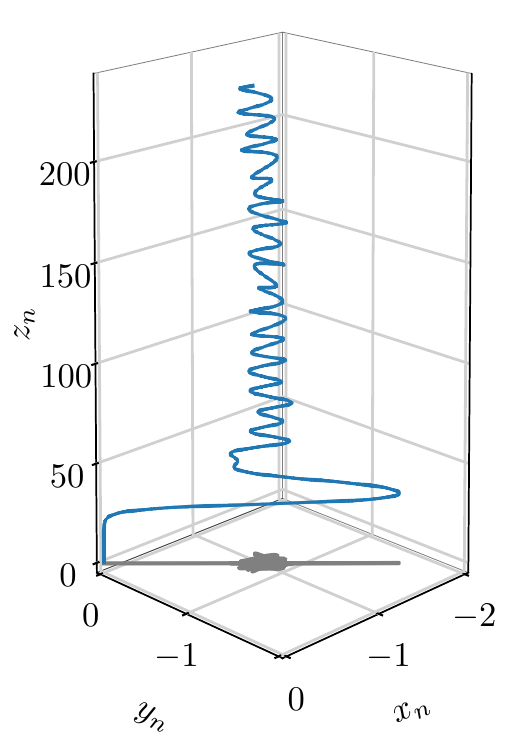}}
	\hfill
	\hfill
	\caption{Cases 5, 10 and 11: Various cases of zig-zagging in their vertical and horizontal (projected to xy plane) movement.}
\end{figure}

Various trajectory types are observed at a Galilei number of 200 and three different density ratios. 
Planar and regular zig-zagging occurs for Case 5 at $\pip = 0.5$, as can be seen in \cref{fig:rising_sphere_180_ga_300_all_xyz,fig:rising_sphere_ga_200_pi-0.5_xy}, where the lateral displacement during each ZZ iteration measures about 1.5$d_p$. 
The Strouhal number is $\St = 0.025$.
\AM~did not state detailed numbers for this set of parameters,
yet depicted the corresponding path, which matches with ours in \cref{fig:rising_sphere_180_ga_300_all_xyz} (blue).
\ZD~more extensively reported on this case, providing a frequency
spectrum of the horizontal velocity. 
They noticed a main peak at $f_n = 0.035$, which coincides with ours very well ($0.035$). 
The same can be stated about a second, smaller peak at $f_n = 0.10$. 
Furthermore, the descriptions of the particle motion correspond well.

Case 6 lowers the density ratio to $0.1$, yielding
an irregular zig-zagging pattern
with some obliqueness up to $z_n \approx 100$ (\cref{fig:rising_sphere_180_ga_300_all_xyz}). 
Also, three-dimensional motion
up to a certain degree is observed. 
As such, no Strouhal number can be determined. 
This differs from \AM, who report periodic oblique paths here. 
Only the initial phase up to $z_n \approx 70$ agrees qualitatively with their simulations.

Similar observations are made for Case 7 at $\pip = 0.05$ in \cref{fig:rising_sphere_ga_200_pi_0.05_and_0.001_xyz},
which starts with a regular PO movement, continues obliquely with reduced periodicity and ends chaotic after rising up to $z_n \approx 160$.
Up until that point, the average inclination of $2.9\degree$ falls within the range of $< 3\degree$ reported by \AM.

In Case 8, a density ratio of $\pip = 0.001$ is simulated by utilizing the \vmmem. 
Initially, the path resembles the one of the previous Case 7, but stays approximately planar. 
Additionally, its average inclination of $3 \degree$ does not change much during 
its 
periodic zig-zags, such that this case can be assigned to the PO regime. 
As such, a Strouhal number of $0.045$ for the dominant 
frequency ($f_n = 0.063$) can be determined. 
Worth mentioning is the emergence of a second not as pronounced frequency at $f_n = 0.111$ with $\St = 0.079$. 
While \ZD~did not provide detailed results on the last three cases,
their provided classification suggests chaotic movement for $\Ga = 200$ and $\pip \leq 0.2$.

\begin{figure}
		\centering
		\includegraphics[width=.75\textwidth]{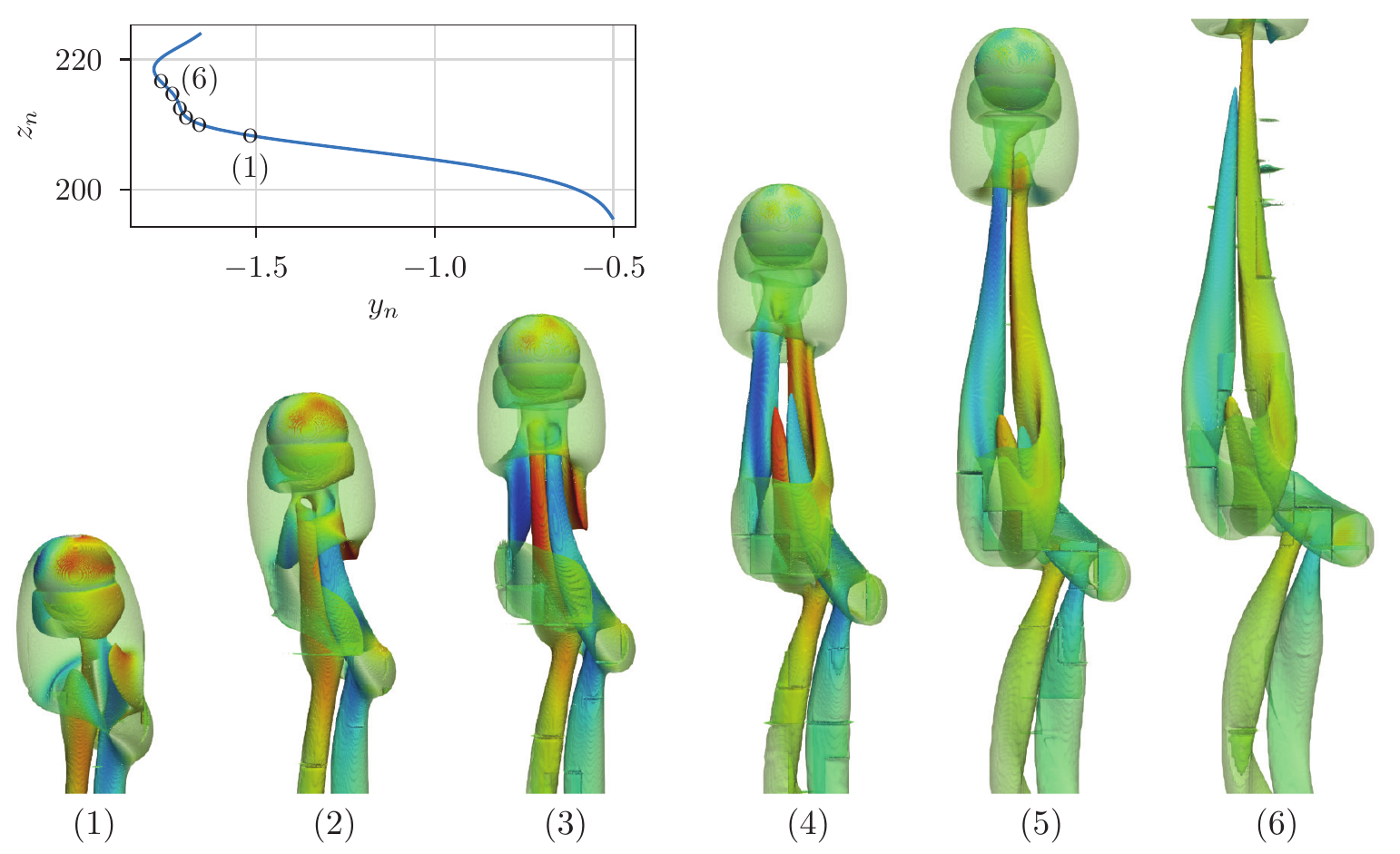}
		\caption{Case 9: Vortices behind a sphere of density ratio $0.5$ at $\Ga = 212.5$. Depicted are the states before and during movement through a crest of zig-zagging. A Q criterion threshold of $10^{-10}$ is employed and coloring is done by normalized vorticity, which ranges between -4 (blue) and 4 (red), green corresponding to 0.}
		\label{fig:ga_212_wake}
\end{figure}

Case 9 is displayed in \cref{fig:ga_212_wake}, where 
the sphere's motion at Galilei number of $212.5$ and density ratio of $0.5$ behaves comparable to Case 5.
It corresponds to the ZZ regime, slightly tilting its original zig-zagging plane beyond $z_n \approx 180$. 
The projected horizontal displacement of the zig-zagging of approximately $1.5d_p$ is also similar. 
\AM~identified a Strouhal number ranging from $0.016$ to $0.036$, confirming ours at $\St \approx 0.028$. 

\cref{fig:rising_sphere_ga_237.5_pi-0.5_xy} illustrates the trajectory of 
a particle at $\Ga = 237.5$ and $\pip = 0.5$, corresponding to Case 10. 
The path starts as ZZ until $z_n$ reaches 100 with a Strouhal number of $\St = 0.03$, 
having a large amplitude. 
Afterwards it enters a slightly inclined plane ($1.37\degree$) with the periodic frequency increasing strongly, 
leading to $\St \approx 0.10$. 
This is different from the findings of \AM, displaying a movement
more prevalent for them at higher Galilei numbers between 250 and 300. 
At this point, the sphere possibly already enters their ZZ$_2$ regime, causing a much larger Strouhal number. 
When considering the simulation at $\Ga = 240$ and $\pip = 0.5$ by \ZD~as reference, 
their results match with ours with the exception of initial large-amplitude zigzagging, as described.

ZZ$_2$ is also the regime in which Case 11 with a 
Galilei number of 300 and density ratio of 0.5 is placed. 
Its spatial progression is depicted in \cref{fig:rising_sphere_ga_300_pi-0.5_xy}. 
Different from \AM's simulation,
we observe an initial lateral swing of magnitude  $2d_p$, before a more regular oscillatory path is taken. 
Furthermore, their Strouhal number approximately coincides with ours, at $\St = 0.10$. 
The distance from crest-to-crest is $0.3d_p$, as in \AM. 
More so, their observation of a rotating zig-zag plane can be confirmed (see $xy$ plot in \cref{fig:rising_sphere_ga_300_pi-0.5_xy}). 
The findings of \ZD~also agree with ours.

\subsubsection{Galilei Numbers $\Ga \geq 400$: Spiralling and Three Dimensionally Chaotic}

The last simulations cover Galilei numbers of 400 and 500, corresponding to Cases 12 to 16. 
Depending on the density ratio, we encounter either chaotic movement or a highly regular spiraling regime.

\begin{figure}
	\captionsetup[subfigure]{position=b}
	\centering
	\hfill
	\subcaptionbox{Case 12: $\Ga = 400, \pip = 0.01$: SP.\label{fig:rising_sphere_ga_400_pi-0.5}}[0.3\linewidth]{\includegraphics[width=0.8\linewidth]{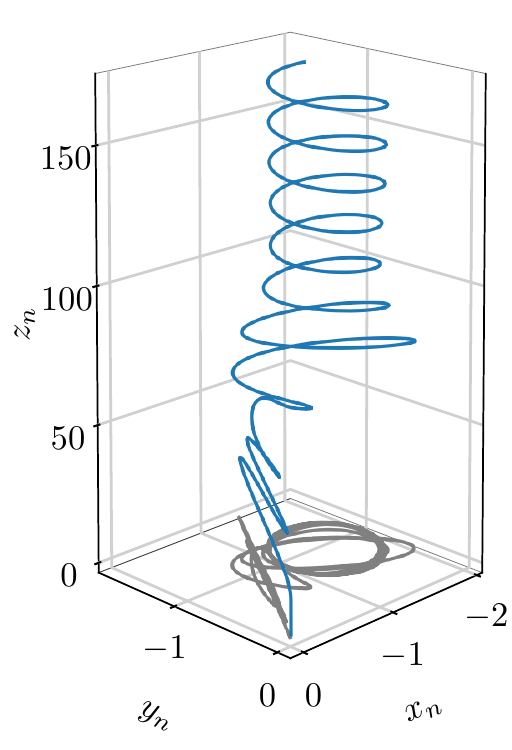}}
	\hfill
	\subcaptionbox{Case 14: $\Ga = 500, \pip = 0.5$: 3DC.\label{fig:rising_sphere_ga_500_pi-0.5}}[0.3\linewidth]{\includegraphics[width=0.8\linewidth]{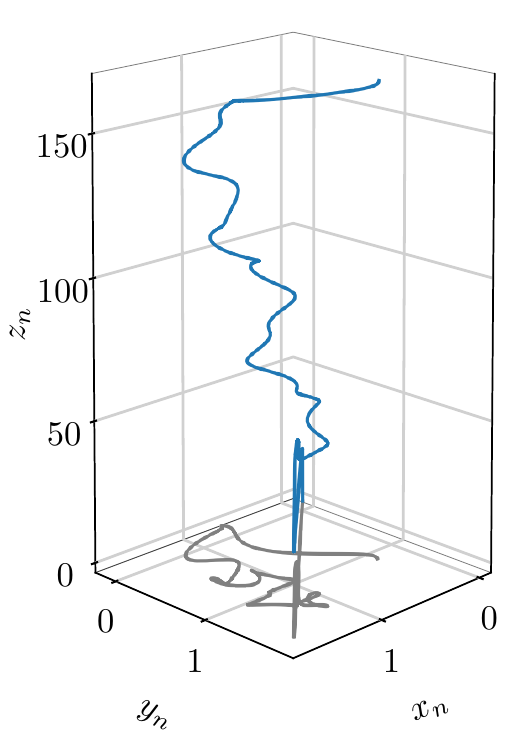}}
	\hfill
	\subcaptionbox{Case 16: $\Ga = 500, \pip = 0.001$: SP.\label{fig:rising_sphere_ga_500_pi-0.001}}[0.3\linewidth]{\includegraphics[width=0.8\linewidth]{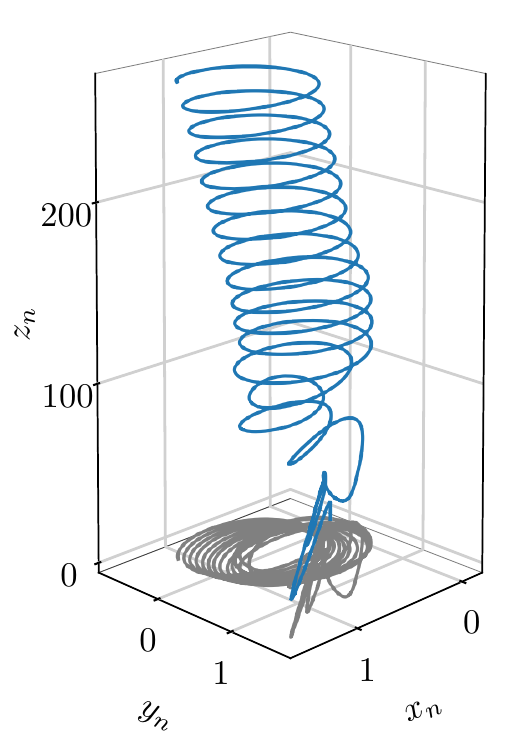}}
	\hfill
	\hfill
	\caption{Cases 12, 14 and 16: At $\Ga = 400$ and above, two regimes are found: Three dimensional chaos (3DC) and spiraling (SP).}
\end{figure}

This new regime is found for Case 12 at $\Ga = 400$ and $\pip = 0.01$.
The spiraling begins after an initial phase, which transforms from planar zig-zagging up until $t_n = 40$ to a three dimensional movement. 
After $t_n = 80$ the projection of the trajectory onto the $xy$ plane resembles an ellipsis closely approaching the shape of a circle with a diameter of $1d_p$. 
The vertical stride between the spiraling iterations is approximately $14.2d_p$, resulting in $\St \approx 0.07$. 
While not matching perfectly with \AM, these numbers come close to their spiral diameter $1.3d_p$ and Strouhal number of approximately $0.07$ ($13.5d_p$ stride). 

Case 13 decreases the particle density ratio further to $0.001$ at $\Ga = 400$ with the aid of the virtual mass approach.
The vertical crest-to-crest distance changes to $14.06d_p$, while the Strouhal number stays the same. 
Equally, the diameter of the spiral varies only to a very minor degree.  
\ZD~provided results of an approximately massless sphere at this Galilei number. 
Its trajectories match ours, however pitch ($12.6d_p$ theirs, $14.06d_p$ ours) and terminal vertical velocity ($1.382$ theirs, $1.466$ ours) differ to a certain degree. 

Continuing with Case 14, we observe the emergence of a three dimensional chaotic region, simulated here using a Galilei number of $500$ and density ratio of $0.5$. 
This path is highly irregular and shows no sign of periodicity (\cref{fig:rising_sphere_ga_500_pi-0.5}). 
The occurrence of 3DC trajectories at this parameter combination was also observed by \AM~and \ZD. 

Case 15 features a particle of density ratio $0.05$ and Galilei number $500$. 
Again, the particle takes up a SP path, albeit with an increased duration of the initial transition compared to the particle of Cases 12 and 13 at $\Ga = 400$. 
Even though at $\pip = 0.001$, \cref{fig:rising_sphere_ga_500_pi-0.001} shows approximately the same delay as $\pip = 0.05$.
The vertical stride of the spiral reduces slightly to $13.8d_p$ at a horizontal displacement of $1d_p$, which is reflected in an increased Strouhal number $\St = 0.067$. 
In this instance, no reference data is available for a direct comparison. 
\ZD~performed two simulations at the same Galilei number and density ratios of $0.1$ and $0$; their depicted trajectory of $\pip = 0.1$ reflected the notion of the present transient from planar zig-zagging to a spiralling path. 
Their resulting Reynolds number of $750$ confirms ours at $\Re \approx 757$. 
After entering the quasi-steady state, the particle of Case 15 moves vertically with $u_{z,n} = 1.512$, which places close to \ZD's at $1.406$ ($\pip = 0$) and $1.501$ ($\pip = 0.1$). 
Radius and pitch of the trajectory closely approach \ZD's as well.

Case 16 displays a modification of the SP trajectory (\cref{fig:rising_sphere_ga_500_pi-0.001}), where the particle at $\pip = 0.001$ and $\Ga = 500$ takes up a slightly inclined spiraling path (inclination angle $\approx 0.2 \degree$). 
Such behavior is also observed by \AM, who found the corresponding motion already at $\pip = 0.01$. 
The spiral is described by a diameter of $1.09d_p$ and vertical stride of $13.22d_p$; the movement along this spiral by $\St = 0.074$ and $u_{z,t} = 1.455$. 

\begin{figure}
		\centering
		\includegraphics[width=0.75\textwidth]{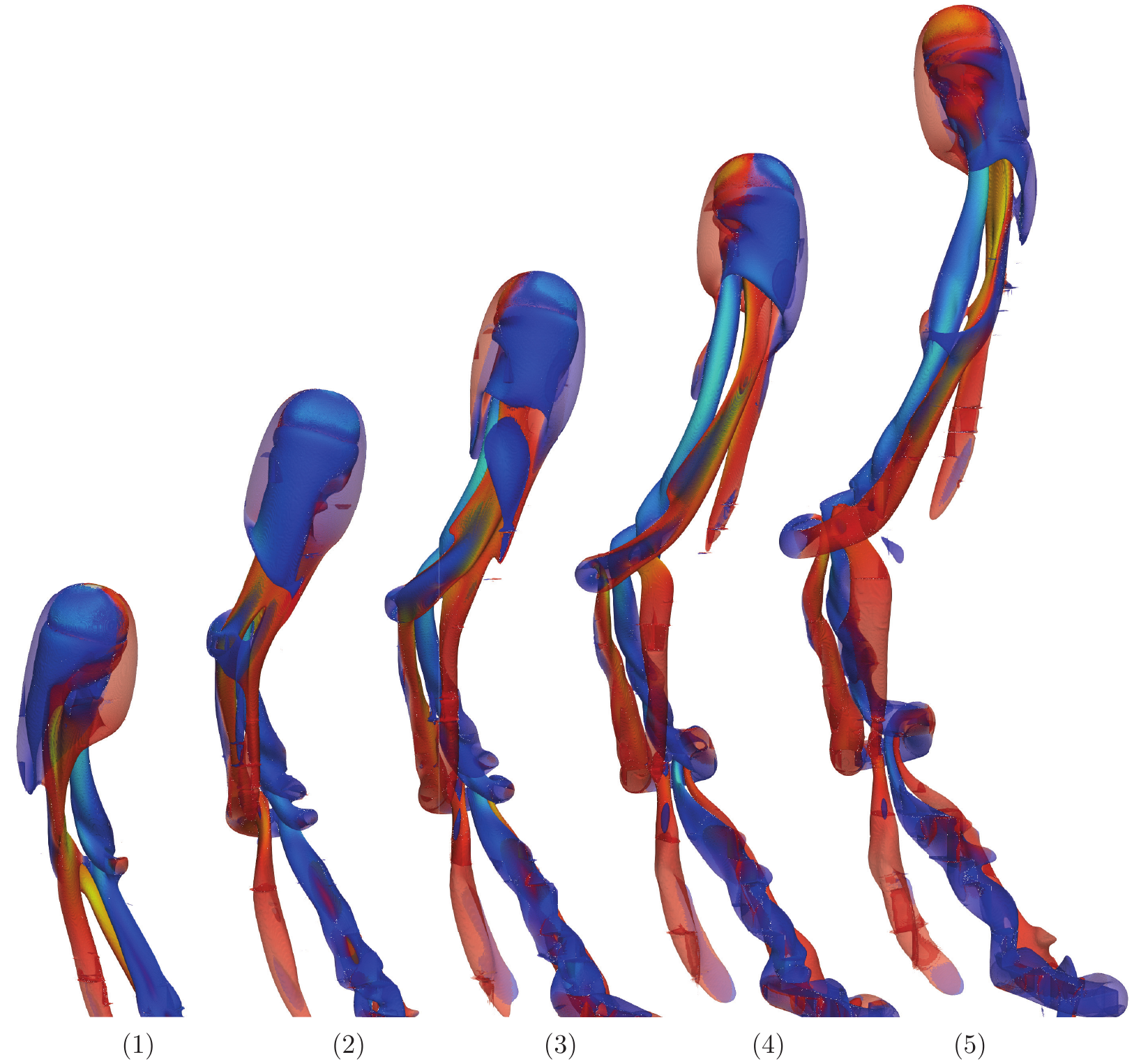}
		\caption{Case 15: Vortices behind a sphere of density ratio $0.05$ at $\Ga = 500$. In frames 4 and 5 the three vortex threads delivering a constant torque to the ascending sphere are well visible. A Q criterion threshold of $10^{-10}$ is employed and coloring is done by normalized vorticity, which ranges between -10 (blue) and 10 (red). Lighter colors correspond to stronger magnitude.}
		\label{fig:ga_500_wake}
\end{figure}

\section{Conclusion}
\label{sec:conclusion}

In this work, we present an 
improved LBM for simulations of 
light particles submerged in a fluid, 
including a comparison with previously established methods.
This is based on a set of benchmarks that are chosen to expose
specific difficulties and peculiarities
arising in systems of submerged particles 
with density ratios $\ll 1$. 
In order to achieve stable simulations without excessively fine resolution,
we adapt and apply the virtual mass approach 
of Schwarz \etal \cite{Schwarz2015}, resulting in the improved \vmmem.
The underlying idea is to artificially increase the mass of the particle to avoid an otherwise vansihing denominator that would amplify inaccuracies of the fluid-particle coupling scheme.
This is exactly compensated by an appropriate force and torque, which in practice requires an approximation of the spheres translational and rotational acceleration.
This approach is shown to enable density ratios of 0.001 and which would thus permit 
e.g.\ the simulation of spherical air bubbles
in water. 
The numerical stabilization scheme is validated both 
with respect to the accuracy of rotational and translational velocities.

In order to further increase the computational efficiency of the 
parallel LBM code, we employ adaptive grid refinement. 
This ensures an adequate and accurate representation of 
the flow features while permitting large computational domains. 
Areas in need of refinement are identified by employing sensors evaluating the state of the fluid 
and the solid phase.
The evaluation of the grid adaptation criteria is invoked at regular time intervals. 
For the case of a rising sphere, 
the computational cost could be reduced by a factor of 71, 
when compared to executing same simulation scenario on a uniform fine grid. 
This is achieved while the accuracy is essentially unaffected.

The efficiency improvement supplied by the adaptive grid refinement algorithm
is finally applied in 16 distinct simulation scenarios that study the trajectories of a single rising sphere. 
In the cases of smallest density ratio of $0.001$, the virtual mass approach proved to be essential to reach
a stable time stepping.
A multitude of 
different trajectory types is observed and
the results are compared to data from the literature.
This includes trajectory classes such as 
vertical to oblique, zig-zagging, spiraling, three dimensional chaos, and intermittent motion.
In many cases good agreement to experimental and numerical studies could be found, for the broad classification via the regimes but also for specific parameters like terminal rising velocities and oscillation frequencies.
This served as a further validation of the here presented approach and showcased its applicability for predictions of complex particle dynamics.
In some cases, the existing literature provided contradicting statements regarding the observed regime of motion and we observed supporting arguments for one or the other.

We note that while validated here for single spherical particles,
our new method is neither restricted to a single particle nor to spherical shapes.
The \vmmem{} is also suitable for a wide range of physical parameters, and it can be employed in domains of arbitrary shape.
This is considered a major improvement to the numerical schemes applied previously in the simulation studies of trajectories of rising particles.
A massively parallel and efficient implementation
is available in the open-source \walberla{} framework\cite{Bauer2020}, 
enabling the simulation of ensembles of several submerged particles and to study their collective motion.
Such studies will be the topic of future research.

\section*{Acknowledgments}

The authors gratefully acknowledge the Erlangen Regional Computing Center (\url{www.rrze.fau.de}) as well as the Gauss Centre for Supercomputing e.V. (\url{www.gauss-centre.eu}) for funding this project by providing computing time on their supercomputers.

\bibliography{library}%

\begin{thebibliography}{10}
\providecommand \doibase [0]{http://dx.doi.org/}%

\bibitem{Andrady2011}
Andrady AL. Microplastics in the marine environment. {\it Marine Pollution
  Bulletin} 2011\string; 62(8)\string: 1596 - 1605.
\newblock \href {\doibase https://doi.org/10.1016/j.marpolbul.2011.05.030}
  {doi: https://doi.org/10.1016/j.marpolbul.2011.05.030}

\bibitem{Wright2013}
Wright SL, Thompson RC, Galloway TS. {The physical impacts of microplastics on
  marine organisms: a review.}. {\it Environmental pollution (Barking, Essex :
  1987)} 2013\string; 178\string: 483--492.
\newblock \href {\doibase 10.1016/j.envpol.2013.02.031} {doi:
  10.1016/j.envpol.2013.02.031}

\bibitem{Cole2011}
Cole M, Lindeque P, Halsband C, Galloway TS. {Microplastics as contaminants in
  the marine environment: A review}. {\it Marine Pollution Bulletin}
  2011\string; 62(12)\string: 2588--2597.
\newblock \href {\doibase 10.1016/j.marpolbul.2011.09.025} {doi:
  10.1016/j.marpolbul.2011.09.025}

\bibitem{Driedger2015}
Driedger AG, D{\"{u}}rr HH, Mitchell K, {Van Cappellen} P. {Plastic debris in
  the Laurentian Great Lakes: A review}. {\it Journal of Great Lakes Research}
  2015\string; 41(1)\string: 9--19.
\newblock \href {\doibase 10.1016/j.jglr.2014.12.020} {doi:
  10.1016/j.jglr.2014.12.020}

\bibitem{Thorpe1988}
Thorpe SA, Hall AJ. {Bubble clouds and temperature anomalies in the upper
  ocean}. {\it Nature} 1988\string; 328(6125)\string: 48--51.
\newblock \href {\doibase 10.1038/328048a0} {doi: 10.1038/328048a0}

\bibitem{Weller1990}
Weller RA. {Not so quiet on the ocean front}. {\it Nature} 1990\string;
  348(6298)\string: 199--200.
\newblock \href {\doibase 10.1038/348199a0} {doi: 10.1038/348199a0}

\bibitem{Almeras2015}
Alm{\'{e}}ras E, Risso F, Roig V, Cazin S, Plais C, Augier F. {Mixing by
  bubble-induced turbulence}. {\it Journal of Fluid Mechanics} 2015\string;
  776\string: 458--474.
\newblock \href {\doibase 10.1017/jfm.2015.338} {doi: 10.1017/jfm.2015.338}

\bibitem{Mathai2018a}
Mathai V, Huisman SG, Sun C, Lohse D, Bourgoin M. {Enhanced dispersion of big
  bubbles in turbulence}.  2018(1)\string: 1--6.

\bibitem{Mathai2020}
Mathai V, Lohse D, Sun C. {Bubbly and Buoyant Particle-Laden Turbulent Flows}.
  {\it Annual Review of Condensed Matter Physics} 2020\string; 11(1)\string:
  529--559.
\newblock \href {\doibase 10.1146/annurev-conmatphys-031119-050637} {doi:
  10.1146/annurev-conmatphys-031119-050637}

\bibitem{Bourgoin2014}
Bourgoin M, Xu H. {Focus on dynamics of particles in turbulence}. {\it New
  Journal of Physics} 2014\string; 16.
\newblock \href {\doibase 10.1088/1367-2630/16/8/085010} {doi:
  10.1088/1367-2630/16/8/085010}

\bibitem{Mathai2018}
Mathai V, Zhu X, Sun C, Lohse D. {Flutter to tumble transition of buoyant
  spheres triggered by rotational inertia changes}. {\it Nature Communications}
  2018\string; 9(1)\string: 1--7.
\newblock \href {\doibase 10.1038/s41467-018-04177-w} {doi:
  10.1038/s41467-018-04177-w}

\bibitem{Murrow1964}
Murrow H, Henry R. {Self-Induced Balloon Motions}. {\it Journal of Applied
  Meteorology and Climatology} 1964\string; 4(1)\string: 131--138.
\newblock \href {\doibase 10.1175/1520-0450(1965)004<0131:SIBM>2.0.CO;2} {doi:
  10.1175/1520-0450(1965)004<0131:SIBM>2.0.CO;2}

\bibitem{Scoggins1964}
Scoggins JR. {Aerodynamics of spherical balloon wind sensors}. {\it Journal of
  Geophysical Research} 1964\string; 69(4)\string: 591--598.
\newblock \href {\doibase 10.1029/jz069i004p00591} {doi:
  10.1029/jz069i004p00591}

\bibitem{Lugt1983}
Lugt H. {Autorotation}. {\it Annual Review of Fluid Mechanics} 1983\string;
  15\string: 123--147.
\newblock \href {\doibase 10.1146/annurev.fl.15.010183.001011} {doi:
  10.1146/annurev.fl.15.010183.001011}

\bibitem{Jenny2004}
Jenny M, Du{\v{s}}ek J, Bouchet G. {Instabilities and transition of a sphere
  falling or ascending freely in a Newtonian fluid}. {\it Journal of Fluid
  Mechanics} 2004\string; 508(508)\string: 201--239.
\newblock \href {\doibase 10.1017/S0022112004009164} {doi:
  10.1017/S0022112004009164}

\bibitem{Biesheuvel2007}
Biesheuvel A, Veldhuis C. {An experimental study of the regimes of motion of
  spheres falling or ascending freely in a Newtonian fluid}. {\it International
  Journal of Multiphase Flow} 2007\string; 33(10)\string: 1074--1087.
\newblock \href {\doibase 10.1016/j.ijmultiphaseflow.2007.05.002} {doi:
  10.1016/j.ijmultiphaseflow.2007.05.002}

\bibitem{Horowitz2010}
Horowitz M, Williamson CH. {The effect of Reynolds number on the dynamics and
  wakes of freely rising and falling spheres}. {\it Journal of Fluid Mechanics}
  2010\string; 651\string: 251--294.
\newblock \href {\doibase 10.1017/S0022112009993934} {doi:
  10.1017/S0022112009993934}

\bibitem{Ostmann2017}
Ostmann S, Chaves H, Br{\"{u}}cker C. {Path instabilities of light particles
  rising in a liquid with background rotation}. {\it Journal of Fluids and
  Structures} 2017\string; 70\string: 403--416.
\newblock \href {\doibase 10.1016/j.jfluidstructs.2017.02.007} {doi:
  10.1016/j.jfluidstructs.2017.02.007}

\bibitem{Zhou2015}
Zhou W, Du{\v{s}}ek J. {Chaotic states and order in the chaos of the paths of
  freely falling and ascending spheres}. {\it International Journal of
  Multiphase Flow} 2015\string; 75\string: 205--223.
\newblock \href {\doibase 10.1016/j.ijmultiphaseflow.2015.05.010} {doi:
  10.1016/j.ijmultiphaseflow.2015.05.010}

\bibitem{Auguste2018}
Auguste F, Magnaudet J. {Path oscillations and enhanced drag of light rising
  spheres}. {\it Journal of Fluid Mechanics} 2018\string; 841\string: 228--266.
\newblock \href {\doibase 10.1017/jfm.2018.100} {doi: 10.1017/jfm.2018.100}

\bibitem{versteeg2007introduction}
Versteeg HK, Malalasekera W. {\it An introduction to computational fluid
  dynamics: the finite volume method}.
\newblock Pearson education .
\newblock 2007.

\bibitem{moukalled2016finite}
{F. Moukalled, L. Mangani} MD. {\it {The Finite Volume Method in Computational
  Fluid Dynamics}}. 113.
\newblock Springer .
\newblock 2015.

\bibitem{Chen1998}
Chen S, Doolen GD. {Lattice Boltzmann Method for Fluid Flows}. {\it Annual
  Review of Fluid Mechanics} 1998\string; 30(1)\string: 329--364.
\newblock \href {\doibase 10.1146/annurev.fluid.30.1.329} {doi:
  10.1146/annurev.fluid.30.1.329}

\bibitem{Aidun2010}
Aidun CK, Clausen JR. {Lattice-boltzmann method for complex flows}. {\it Annual
  Review of Fluid Mechanics} 2010\string; 42\string: 439--472.
\newblock \href {\doibase 10.1146/annurev-fluid-121108-145519} {doi:
  10.1146/annurev-fluid-121108-145519}

\bibitem{Clift2008}
Clift R, Grace JR, Weber M. {\it {Bubbles, Drops, and Particles}} .
\newblock 1978.

\bibitem{Rettinger2017Riverbed}
Rettinger C, Godenschwager C, Eibl S, et al. Fully Resolved Simulations of Dune
  Formation in Riverbeds. In:  Kunkel JM, Yokota R, Balaji P, Keyes D.
  \kern-2pt, eds. {\it High Performance Computing}Springer International
  Publishing; 2017; Cham\string: 3--21

\bibitem{vowinckel2019}
Vowinckel B, Biegert E, Luzzatto-Fegiz P, Meiburg E. Consolidation of freshly
  deposited cohesive and noncohesive sediment: Particle-resolved simulations.
  {\it Phys. Rev. Fluids} 2019\string; 4\string: 074305.
\newblock \href {\doibase 10.1103/PhysRevFluids.4.074305} {doi:
  10.1103/PhysRevFluids.4.074305}

\bibitem{peng2019}
Peng C, Ayala OM, Wang LP. A direct numerical investigation of two-way
  interactions in a particle-laden turbulent channel flow. {\it Journal of
  Fluid Mechanics} 2019\string; 875\string: 1096–1144.
\newblock \href {\doibase 10.1017/jfm.2019.509} {doi: 10.1017/jfm.2019.509}

\bibitem{benseghier2020}
Benseghier Z, Cuéllar P, Luu LH, Bonelli S, Philippe P. A parallel GPU-based
  computational framework for the micromechanical analysis of geotechnical and
  erosion problems. {\it Computers and Geotechnics} 2020\string; 120\string:
  103404.
\newblock \href {\doibase 10.1016/j.compgeo.2019.103404} {doi:
  10.1016/j.compgeo.2019.103404}

\bibitem{Goetz2010}
Götz J, Iglberger K, Feichtinger C, Donath S, Rüde U. Coupling multibody
  dynamics and computational fluid dynamics on 8192 processor cores. {\it
  Parallel Computing} 2010\string; 36(2)\string: 142 - 151.
\newblock \href {\doibase 10.1016/j.parco.2010.01.005} {doi:
  10.1016/j.parco.2010.01.005}

\bibitem{hasert2014}
Hasert M, Masilamani K, Zimny S, et al. Complex fluid simulations with the
  parallel tree-based Lattice Boltzmann solver Musubi. {\it Journal of
  Computational Science} 2014\string; 5(5)\string: 784 - 794.
\newblock \href {\doibase 10.1016/j.jocs.2013.11.001} {doi:
  10.1016/j.jocs.2013.11.001}

\bibitem{Bauer2020}
Bauer M, Eibl S, Godenschwager C, et al. {WALBERLA: A block-structured
  high-performance framework for multiphysics simulations}. {\it Computers and
  Mathematics with Applications} 2020.
\newblock \href {\doibase 10.1016/j.camwa.2020.01.007} {doi:
  10.1016/j.camwa.2020.01.007}

\bibitem{Ladd1994}
Ladd AJ. {Numerical Simulations of Particulate Suspensions Via a Discretized
  Boltzmann Equation. Part 1. Theoretical Foundation}. {\it Journal of Fluid
  Mechanics} 1994\string; 271\string: 285--309.
\newblock \href {\doibase 10.1017/S0022112094001771} {doi:
  10.1017/S0022112094001771}

\bibitem{Aidun1998}
Aidun CK, Lu Y, Ding EJ. {Direct analysis of particulate suspensions with
  inertia using the discrete Boltzmann equation}. {\it Journal of Fluid
  Mechanics} 1998\string; 373\string: 287--311.
\newblock \href {\doibase 10.1017/S0022112098002493} {doi:
  10.1017/S0022112098002493}

\bibitem{Rettinger2017}
Rettinger C, R{\"{u}}de U. {A comparative study of fluid-particle coupling
  methods for fully resolved lattice Boltzmann simulations}. {\it Computers and
  Fluids} 2017\string; 154\string: 74--89.
\newblock \href {\doibase 10.1016/j.compfluid.2017.05.033} {doi:
  10.1016/j.compfluid.2017.05.033}

\bibitem{Ladd1994_2}
Ladd AJC. Numerical simulations of particulate suspensions via a discretized
  Boltzmann equation. Part 2. Numerical results. {\it Journal of Fluid
  Mechanics} 1994\string; 271\string: 311–339.
\newblock \href {\doibase 10.1017/S0022112094001783} {doi:
  10.1017/S0022112094001783}

\bibitem{Uhlmann2005}
Uhlmann M. {An immersed boundary method with direct forcing for the simulation
  of particulate flows}. {\it Journal of Computational Physics} 2005\string;
  209(2)\string: 448--476.
\newblock \href {\doibase 10.1016/j.jcp.2005.03.017} {doi:
  10.1016/j.jcp.2005.03.017}

\bibitem{Kempe2012}
Kempe T, Fr{\"{o}}hlich J. {An improved immersed boundary method with direct
  forcing for the simulation of particle laden flows}. {\it Journal of
  Computational Physics} 2012.
\newblock \href {\doibase 10.1016/j.jcp.2012.01.021} {doi:
  10.1016/j.jcp.2012.01.021}

\bibitem{Breugem2012}
Breugem WP. {A second-order accurate immersed boundary method for fully
  resolved simulations of particle-laden flows}. {\it Journal of Computational
  Physics} 2012\string; 231(13)\string: 4469--4498.
\newblock \href {\doibase 10.1016/j.jcp.2012.02.026} {doi:
  10.1016/j.jcp.2012.02.026}

\bibitem{Inamuro2004}
Inamuro T, Ogata T, Tajima S, Konishi N. {A lattice Boltzmann method for
  incompressible two-phase flows with large density differences}. {\it Journal
  of Computational Physics} 2004\string; 198(2)\string: 628--644.
\newblock \href {\doibase 10.1016/j.jcp.2004.01.019} {doi:
  10.1016/j.jcp.2004.01.019}

\bibitem{Apte2013}
Apte SV, Finn JR. {A variable-density fictitious domain method for particulate
  flows with broad range of particle-fluid density ratios}. {\it Journal of
  Computational Physics} 2013\string; 243\string: 109--129.
\newblock \href {\doibase 10.1016/j.jcp.2012.12.021} {doi:
  10.1016/j.jcp.2012.12.021}

\bibitem{banks2018}
Banks J, Henshaw W, Schwendeman D, Tang Q. A stable partitioned FSI algorithm
  for rigid bodies and incompressible flow in three dimensions. {\it Journal of
  Computational Physics} 2018\string; 373\string: 455-492.
\newblock \href {\doibase 10.1016/j.jcp.2018.06.072} {doi:
  10.1016/j.jcp.2018.06.072}

\bibitem{Jenny2004a}
Jenny M, Du{\v{s}}ek J. {Efficient numerical method for the direct numerical
  simulation of the flow past a single light moving spherical body in
  transitional regimes}. {\it Journal of Computational Physics} 2004\string;
  194(1)\string: 215--232.
\newblock \href {\doibase 10.1016/j.jcp.2003.09.004} {doi:
  10.1016/j.jcp.2003.09.004}

\bibitem{Hu2001}
Hu HH, Patankar NA, Zhu MY. {Direct Numerical Simulations of Fluid-Solid
  Systems Using the Arbitrary Lagrangian-Eulerian Technique}. {\it Journal of
  Computational Physics} 2001\string; 169(2)\string: 427--462.
\newblock \href {\doibase 10.1006/jcph.2000.6592} {doi: 10.1006/jcph.2000.6592}

\bibitem{Schwarz2015}
Schwarz S, Kempe T, Fr{\"{o}}hlich J. {A temporal discretization scheme to
  compute the motion of light particles in viscous flows by an immersed
  boundary method}. {\it Journal of Computational Physics} 2015\string;
  281\string: 591--613.
\newblock \href {\doibase 10.1016/j.jcp.2014.10.039} {doi:
  10.1016/j.jcp.2014.10.039}

\bibitem{Rettinger2020b}
Rettinger C, R{\"{u}}de U. {An efficient four-way coupled lattice Boltzmann -
  discrete element method for fully resolved simulations of particle-laden
  flows}.  2020\string: 1--37.

\bibitem{krueger2017}
Kr{\"u}ger T, Kusumaatmaja H, Kuzmin A, Shardt O, Silva G, Viggen EM. {\it The
  lattice {B}oltzmann method}.
\newblock Springer .
\newblock 2017

\bibitem{Qian1992}
Qian YH, D'Humi{\`{e}}res D, Lallemand P. {Lattice bgk models for navier-stokes
  equation}. {\it Epl} 1992\string; 17(6)\string: 479--484.
\newblock \href {\doibase 10.1209/0295-5075/17/6/001} {doi:
  10.1209/0295-5075/17/6/001}

\bibitem{he1997}
He X, Luo LS. Lattice {Boltzmann} model for the incompressible
  {Navier}--{Stokes} equation. {\it Journal of Statistical Physics}
  1997\string; 88(3-4)\string: 927--944.
\newblock \href {\doibase 10.1023/B:JOSS.0000015179.12689.e4} {doi:
  10.1023/B:JOSS.0000015179.12689.e4}

\bibitem{DHumieres2002}
D'Humi{\`{e}}res D, Ginzburg I, Krafczyk M, Lallemand P, Luo LS.
  {Multiple-relaxation-time lattice Boltzmann models in three dimensions}. {\it
  Philosophical Transactions of the Royal Society A: Mathematical, Physical and
  Engineering Sciences} 2002\string; 360(1792)\string: 437--451.
\newblock \href {\doibase 10.1098/rsta.2001.0955} {doi: 10.1098/rsta.2001.0955}

\bibitem{duenweg2007}
D\"unweg B, Schiller UD, Ladd AJC. Statistical mechanics of the fluctuating
  lattice {B}oltzmann equation. {\it Phys. Rev. E} 2007\string; 76\string:
  036704.
\newblock \href {\doibase 10.1103/PhysRevE.76.036704} {doi:
  10.1103/PhysRevE.76.036704}

\bibitem{Ginzburg2008}
Ginzburg I, Verhaeghe F, D'Humi{\`{e}}res D. {Two-relaxation-time Lattice
  Boltzmann scheme: About parametrization, velocity, pressure and mixed
  boundary conditions}. {\it Communications in Computational Physics}
  2008\string; 3(2)\string: 427--478.

\bibitem{khirevich2015}
Khirevich S, Ginzburg I, Tallarek U. Coarse- and fine-grid numerical behavior
  of MRT/TRT lattice-{B}oltzmann schemes in regular and random sphere packings.
  {\it Journal of Computational Physics} 2015\string; 281\string: 708 - 742.
\newblock \href {\doibase 10.1016/j.jcp.2014.10.038} {doi:
  10.1016/j.jcp.2014.10.038}

\bibitem{Preclik2015}
Preclik T, R{\"{u}}de U. {Ultrascale simulations of non-smooth granular
  dynamics}. {\it Computational Particle Mechanics} 2015\string; 2(2)\string:
  173--196.
\newblock \href {\doibase 10.1007/s40571-015-0047-6} {doi:
  10.1007/s40571-015-0047-6}

\bibitem{Wen2014}
Wen B, Zhang C, Tu Y, Wang C, Fang H. {Galilean invariant fluid-solid
  interfacial dynamics in lattice Boltzmann simulations}. {\it Journal of
  Computational Physics} 2014\string; 266\string: 161--170.
\newblock \href {\doibase 10.1016/j.jcp.2014.02.018} {doi:
  10.1016/j.jcp.2014.02.018}

\bibitem{dorschner2015}
Dorschner B, Chikatamarla S, Bösch F, Karlin I. Grad's approximation for
  moving and stationary walls in entropic lattice {B}oltzmann simulations. {\it
  Journal of Computational Physics} 2015\string; 295\string: 340 - 354.
\newblock \href {\doibase 10.1016/j.jcp.2015.04.017} {doi:
  10.1016/j.jcp.2015.04.017}

\bibitem{Holzer2008}
H{\"{o}}lzer A, Sommerfeld M. {New simple correlation formula for the drag
  coefficient of non-spherical particles}. {\it Powder Technology} 2008\string;
  184(3)\string: 361--365.
\newblock \href {\doibase 10.1016/j.powtec.2007.08.021} {doi:
  10.1016/j.powtec.2007.08.021}

\bibitem{Tavanashad2020}
Tavanashad V, Subramaniam S. Fully resolved simulation of dense suspensions of
  freely evolving buoyant particles using an improved immersed boundary method.
  {\it International Journal of Multiphase Flow} 2020\string; 132\string:
  103396.
\newblock \href {\doibase 10.1016/j.ijmultiphaseflow.2020.103396} {doi:
  10.1016/j.ijmultiphaseflow.2020.103396}

\bibitem{schornbaum2016massively}
Schornbaum F, R\"ude U. Massively parallel algorithms for the lattice Boltzmann
  method on nonuniform grids. {\it SIAM Journal on Scientific Computing}
  2016\string; 38(2)\string: C96--C126.
\newblock \href {\doibase 10.1137/15M1035240} {doi: 10.1137/15M1035240}

\bibitem{Rohde2006}
Rohde M, Kandhai D, Derksen JJ, Akker v.~dHE. {A generic, mass conservative
  local grid refinement technique for lattice-Boltzmann schemes}. {\it
  International Journal for Numerical Methods in Fluids} 2006\string;
  51(4)\string: 439--468.
\newblock \href {\doibase 10.1002/fld.1140} {doi: 10.1002/fld.1140}

\bibitem{schornbaum2018extreme}
Schornbaum F, R\"ude U. Extreme-scale block-structured adaptive mesh
  refinement. {\it SIAM Journal on Scientific Computing} 2018\string;
  40(3)\string: C358--C387.
\newblock \href {\doibase 10.1137/17M1128411} {doi: 10.1137/17M1128411}

\bibitem{Crouse2003}
Crouse B, Rank E, Krafczyk M, T{\"{o}}lke J. {A LB-based approach for adaptive
  flow simulations}. {\it International Journal of Modern Physics B}
  2003\string; 17(1-2)\string: 109--112.
\newblock \href {\doibase 10.1142/s0217979203017163} {doi:
  10.1142/s0217979203017163}

\bibitem{Yuan2018}
Yuan R, Zhong C. {An immersed-boundary method for compressible viscous flows
  and its application in the gas-kinetic BGK scheme}. {\it Applied Mathematical
  Modelling} 2018\string; 55\string: 417--446.
\newblock \href {\doibase 10.1016/j.apm.2017.10.003} {doi:
  10.1016/j.apm.2017.10.003}

\bibitem{Deister1999}
Deister F, Hirschelt EH. {Adaptive cartesian/prism grid generation and
  solutions for arbitrary geometries}. {\it 37th Aerospace Sciences Meeting and
  Exhibit} 1999(c).
\newblock \href {\doibase 10.2514/6.1999-782} {doi: 10.2514/6.1999-782}

\bibitem{Zeeuw1993}
{De Zeeuw} D. {\it {A quadtree-based adaptively-refined Cartesian-grid
  algorithm for solution of the Euler equations}}. PhD thesis.  1993

\bibitem{Hunt1988}
Hunt J, Wray A, Moin P. {Eddies, streams, and convergence zones in turbulent
  flows}. {\it Center for Turbulence Research, Proceedings of the Summer
  Program} 1988(1970)\string: 193--208.

\bibitem{Jenny2003}
Jenny M, Bouchet G, Du{\v{s}}ek J. {Nonvertical ascension or fall of a free
  sphere in a Newtonian fluid}. {\it Physics of Fluids} 2003\string;
  15(1)\string: L9--L12.
\newblock \href {\doibase 10.1063/1.1529179} {doi: 10.1063/1.1529179}

\end{thebibliography}

\end{document}